\pdfoutput=1

\documentclass[fleqn,11pt]{article}
\usepackage[letterpaper,left=0.8in,top=0.8in,bottom=0.9in,right=0.8in]{geometry} 

\usepackage[english]{babel}
\usepackage{amsthm}
\usepackage{amsfonts}
\usepackage{amssymb}
\usepackage{mathtools}
\usepackage{chemformula}
\usepackage[autostyle]{csquotes}
\usepackage{url}
\usepackage{comment}
\usepackage[binary-units=true]{siunitx}
\usepackage{afterpage}
\usepackage[
colorlinks = true,
linkcolor = blue,
urlcolor  = blue,
citecolor = blue,
anchorcolor = blue
]{hyperref}
\usepackage{soul}
\usepackage{booktabs}
\usepackage{enumitem}
\usepackage{caption}

\def\@citecolor{blue}
\def\@urlcolor{blue}%
\def\@linkcolor{blue}%

\def\orcidID#1{\href{http://orcid.org/#1}{\protect\includegraphics{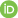}}}
\newcommand{\ECnum}[4]{%
	EC %
	\ifthenelse{\equal{#1}{}}{-.-.-.-}{%
		#1.\ifthenelse{\equal{#2}{}}{-.-.-}{%
			#2.\ifthenelse{\equal{#3}{}}{-.-}{%
				#3.\ifthenelse{\equal{#4}{}}{-}{#4}%
			}%
		}%
	}%
}
\newcommand{\ECnumsimple}[4]{EC #1.#2.#3.#4}  
\newcommand{\uniprot}[1]{UniProtKB: \href{https://www.uniprot.org/uniprot/#1}{#1}}
\newcommand{\mcsaentry}[1]{\href{https://www.ebi.ac.uk/thornton-srv/m-csa/entry/#1/}{entry #1}}

\title{Representing catalytic mechanisms with rule composition}

\author{%
	Jakob L. Andersen,~%
	Rolf Fagerberg,~%
	Christoph Flamm,~%
	Walter Fontana,\\%
	Juri Kol\v{c}\'ak\textsuperscript{$\ast$},~%
	Christophe V.F.P. Laurent,~%
	Daniel Merkle\textsuperscript{$\ast$}~%
	and	Nikolai N{\o}jgaard%
}
\newcommand{\denmark}{%
	Department of Mathematics and Computer Science, University of Southern Denmark, Campusvej 55, 5230 Odense, Denmark%
}
\newcommand{\vienna}{%
	Department of Theoretical Chemistry, University of Vienna, W\"ahringer Stra{\ss}e 17, 1090 Vienna, Austria%
}
\newcommand{\boston}{%
	Department of Systems Biology, Harvard Medical School, 200 Longwood Avenue, 02115 Boston, MA, USA%
}
\newcommand{\acsauthor}[3]{%
	\textbf{#1 --} \textit{#2}; \protect\includegraphics{orcid_color.eps} \href{http://orcid.org/#3}{orcid.org/#3}%
}

\newcommand{\derivation}[3]{#1 \xRightarrow{#3} #2}

\newcommand{\funccomp}{\circ}
\newcommand{\rulecomp}{\bullet}

\newcommand{\ourITSname}{OG}

\begin{document}

\maketitle

\begin{abstract}
\noindent Reaction mechanisms are often presented as sequences of elementary steps, such as codified by arrow pushing. We propose an approach for representing such mechanisms using graph transformation. In this framework, each elementary step is a rule for modifying a molecular graph and a mechanism is a sequence of such rules. To generate a compact representation of a multi-step reaction, we compose the rules of individual steps into a composite rule, providing a rigorous and fully automated approach to coarse-graining. While the composite rule retains the graphical conditions necessary for the execution of a mechanism, it also records information about transient changes not visible by comparing educts and products. By projecting the rule onto a single \enquote{overlay graph}, we generalize Fujita's idea of an Imaginary Transition Structure~\cite{Fujita1986} from elementary reactions to composite reactions. The utility of the overlay graph construct is exemplified in the context of enzyme-catalyzed reactions. In a first application, we exploit mechanistic information in the Mechanism and Catalytic Site Atlas~\cite{RibeiroHFTFT2017} to construct overlay graphs of hydrolase reactions listed in the database. These graphs point at a spectrum of catalytic entanglement of enzyme and substrate, de-emphasizing the notion of a singular catalyst in favor of a collection of catalytic sites that can be distributed across enzyme and substrate. In a second application, we deploy composite rules to search the Rhea database~\cite{LombardotMAAHNIXCRB2019} for reactions of known or unknown mechanism that are, in principle, compatible with the mechanisms implied by the composite rules. We believe this work adds to the utility of graph-transformation formalisms in representing and reasoning about chemistry in an automated yet insightful fashion.
\end{abstract}


\section{Introduction}
\label{sec:introductionAlt}

A chemical reaction is often viewed as occurring in a sequence of steps considered to be elementary at some level of abstraction. One such level is the concerted net movement of one or two electrons from a center high in electronic density to a center low in electronic density, as codified by \enquote{arrow pushing}~\cite{Levy2011}. A mechanism is then a sequence of elementary steps---or steps, for short---that reach a stable configuration. At a higher level of abstraction, one might view a complete reaction as an \enquote{elementary} step and a synthesis path or metabolic pathway as a mechanism.

A mechanism explains the chemistry of a reaction by providing a causal account of how (and why) a particular outcome is attained. Cyclical mechanisms, such as catalysis, are of particular interest in biology and industry. Since the catalyst is recycled, any molecular change it undergoes in one step of the mechanism must be undone in subsequent steps. Much of the chemistry may then consist of transient changes that remain invisible at the granularity of the overall reaction, which adds up all the steps. This is the granularity at which many enzymatic reactions are expressed. Our aim in this paper is to develop a representation of a catalyzed multi-step reaction that is compact like an overall reaction yet richer in that it captures transient modifications. 

Although fairly intuitive to grasp, this representation must be made formal. Formality provides it with clear computational semantics for automatic construction and deployment. We approach this task within the framework of graph transformation~\cite{EhrigEGT2006}. It is standard practice to draw a molecular structure formula as an undirected graph with vertices labeled by atom types and edges by bond types. Since a reaction is the modification of educt graphs into product graphs, methods for expressing graph transformations~\cite{HabelMP2001} provide a natural representation of chemistry. A graph transformation is, roughly, a rule $L \rightarrow R$ with $L$ and $R$ denoting structural patterns. Any molecular structure graph that matches $L$ qualifies for the transformation, which consists in replacing the matching part with $R$. This is akin to what chemists call a reaction template. The difference is that a graph transformation must specify which atoms in $L$ correspond to which atoms in $R$, i.e.~the atom map. It is because of the atom-map requirement that a rule captures an aspect of the mechanism not provided by a reaction template. This requirement is also necessary for the notion of graph transformation to be expressed in a mathematically rigorous manner.

To the best of our knowledge, the currently only open-source graph-transformation platform for chemistry based on an atom-level graph model of molecules is MØD~\cite{AndersenFMS2016}. Given a collection of reactants, MØD identifies the graphs that match a pattern specified by a rule and returns the products that result from applying the rule. In this way, MØD can apply a chemistry, codified as a set of rules, to an initial set of reactants and iteratively determine the implied reaction network to any depth as computational resources permit. 

Unfortunately, atom maps of chemical reactions are not provided as often as one would hope. However, elementary steps or reactions are typically expressed in terms of \enquote{arrow pushing} diagrams, which clearly imply the atom map. Such steps can be translated into graph-transformation rules. A multi-step reaction, such as an enzyme-catalyzed reaction, can then be represented as the sequential application of step-specific rules. The main contribution of the work presented here consists in expressing this sequence succinctly as a single rule obtained by the formal composition of all step-specific rules. It turns out that an equivalent representation of this composite rule is an overlay of so-called imaginary transition structures for each elementary step. The notion of an imaginary transition structure (ITS) was introduced by Fujita~\cite{Fujita1986}. An ITS is a structural formula in which the right-hand side of a reaction is superposed on the left, yielding a classification of each bond according to whether it is present only in the educts, only in the products, or in both. The overlay graph, or OG for short, introduced in the present paper naturally extends the concept of ITS with additional classes of transient changes. Tracking transient changes is particularly important when dealing with catalytic mechanisms, since the catalyst may, and indeed often does, swap atoms with the substrate(s). 

We believe the OG to be a useful conceptual and computational tool for describing, analyzing, searching, and designing catalyzed reactions. Some first use cases in support of this belief are provided. In a first application, we exploit arrow pushing information in the Mechanism and Catalytic Site Atlas~\cite{RibeiroHFTFT2017} (M-CSA) to automatically construct OGs of hydrolase reactions listed in the database. These OGs point at a spectrum of catalytic entanglements of enzyme and substrate, de-emphasizing the notion of a singular catalyst in favor of a collection of catalytic sites that can be distributed across enzyme and substrate. In a second application, we deploy composite rules to automatically search the Rhea database~\cite{LombardotMAAHNIXCRB2019} for reactions of known or unknown mechanisms (i.e.~no atom map) that are compatible with the mechanism captured by a composite rule.

Since the graph-transformation approach can be heavy for those unfamiliar with it, we structured the main text of the paper to be largely devoid of formalism.


\section{Methods}
\label{sec:method}

In this section, we convey the idea of an OG by sketching its construction in an informal manner. A rigorous specification, capable of guiding software implementation, is provided in the Supporting Information at two levels of sophistication---one that is mathematically precise yet still close to chemistry and another that is very formal by deploying the language of category theory, which is the \emph{lingua franca} for those in the mathematical branch of the graph transformation business.

To be meaningfully informal, however, we still need to clarify vocabulary. The term \enquote{rule} is typically used by chemists interchangeably with the notion of a reaction template. The idea of a reaction template, $L \rightarrow R$, is to only write those parts, $L$, of molecular structures that are required for a reaction and to specify that $L$ is to be rearranged into $R$. Some parts of $L$ might be atom groups that enable a chemical transformation (by shaping electronic density or playing a steric role) but are not themselves modified. Chemists represent molecular structures as graphs and a pattern is a molecular graph in which some parts are omitted. In the literature, this omission is typically flagged with some remainder symbol. While a remainder symbol like \ch{-R} is a useful visual cue, it is a distraction here; we simply leave off the parts that we don't care about. To determine whether a template can be \emph{applied}, the reactants must \emph{match} the pattern $L$. By rearranging the matching parts in the reactants according to the right side $R$ of the template we obtain a reaction between fully specified molecular structures. Templates are often applied to a collection of molecules by selecting a combination of reactants that match. We use the term \enquote{mixture} as a shorthand for \enquote{a collection of molecules}. A mixture can be very large, like the contents of a test tube, but it can equally refer to a more abstract situation comprising a few reactants that are all used up in a single template application.

\begin{figure}[!h]
\begin{center}
\includegraphics{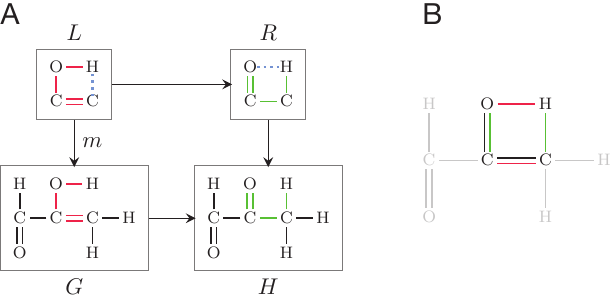}

\caption{{\bf A:} An illustrative example of a rule describing a keto-enol tautomerization, applied to 2-hydroxyacrylaldehyde. The dotted lines represent the non-bond constraint (see text) that the corresponding atoms \emph{not} be bound to each other. For example, on the educt side, the \ch{C} and \ch{H} atoms must not be bound to each other since the rule creates that bond. Given the informal presentation in this section, we only insinuate the existence of an atom map between educt and product molecules by using a spatial layout suggesting which atoms on the left of a rule correspond to which atoms on the right. {\bf B:} The ITS~\cite{Fujita1986} of the reaction in panel A. The ITS is an overlay of the $R$ and $L$ pattern of a rule. In a rendering of the overlay, red and green colors indicate bonds that are broken and created, respectively. The parts that are not taking part in the transformation are faded out in gray. Since the keto-enol tautomerization is an elementary reaction, the ITS is a cycle~\cite{Fujita1986}. See the \hyperref[app:gt_in_chem]{Supporting Information Section~\ref{app:gt_in_chem}} for the case when formal charges are modified.}
\label{fig:rule}

\end{center}
\end{figure}

The key point to note is that this notion of template does not necessarily capture the \emph{mechanism} of a reaction, because it does not require knowing which atoms on the left correspond to which on the right. An atom map is necessary for specifying a physical process of \emph{how} parts are rearranged beyond solely stating their rearrangement. (The atom map is insufficient for specifying the full mechanism since it does not fix the temporal order of physical changes.) While the atom map is not needed for many chemical applications, it plays a crucial role in this paper. Throughout this contribution we therefore reserve the term \enquote{rule} for a reaction template augmented with an atom map. A rule is formally defined in the \hyperref[app:gt_in_chem]{Supporting Information (Section~\ref{app:gt_in_chem})}. On a less formal level, drawings suffice. The top row of~\hyperref[fig:rule]{Figure~\ref{fig:rule}A} illustrates a rule and the bottom row represents its application to a mixture, consisting in this case of a single molecule. The arrow labeled $m$ indicates a matching.

\hyperref[fig:rule]{Figure~\ref{fig:rule}A} points to a subtlety regarding patterns that we need to clarify. The omission of structural parts in forming a pattern means that we do not care what other parts take their place when deciding whether a molecule matches the pattern. However, the omission of a bond between two particular atoms is ambiguous as it could mean that the bond is optional or that its absence is required. For example, the rule in~\hyperref[fig:rule]{Figure~\ref{fig:rule}A} creates a bond between an \ch{H} and a \ch{C} atom and thus requires them \emph{not} to be bound at the outset. We use a dotted line between two specific atoms to assert the required absence of a bond between them. (Similarly, as also illustrated in~\hyperref[fig:rule]{Figure~\ref{fig:rule}A}, breaking a bond between two specific atoms results in a product pattern $R$ in which these atoms are doubtlessly not bound to each other, a fact flagged by a dotted line). Dotted lines never appear in a mixture where molecules are fully specified.

Fujita's ITS, shown in \hyperref[fig:rule]{Figure~\ref{fig:rule}B}, is but an alternative notation for a rule. Instead of using an arrow separating the structure graphs before and after the chemical transformation, the ITS provides instructions for \emph{editing} the reactant graph to obtain the product graph by means of a color-coded (edge symbols in the original ITS) representation of the difference between $R$ and $L$. Obviously, the atom map must be known to edit a molecular structure or pattern. The ITS, therefore, captures the same mechanistic information as a rule.

Obtaining rules is a challenge because the atom map is oftentimes not known and reaction perception is notoriously difficult. Yet, reactions that are considered \enquote{elementary} are typically supported by mechanisms expressed as arrow pushing diagrams, which directly yield the atom map. In this paper, we generate rules from elementary steps of enzymatic mechanisms annotated with arrow pushing diagrams and cataloged in the M-CSA database~\cite{RibeiroHFTFT2017}.
 
\subsection{Overlay Graphs (OGs)}
\label{sec:omg}

Consider a rule $r_1$ that is applied to reactants to generate products that become reactants of a rule $r_2$ yielding further products. We can compress this reaction sequence into an \enquote{overall reaction} that collects all reactants and all products (including those generated by $r_1$ and ignored by $r_2$). The rule capturing this overall reaction is the composition of $r_1$ with $r_2$~\cite{bp2018, AndersenFMS2018}. Rules $r_1$ and $r_2$ each have their atom map, but to compose them we need to track where in the products of $r_1$ there is a match for $r_2$. Multiple matches may exist and so $r_1$ and $r_2$ can give rise to several composite rules, reflecting distinct mechanisms.

\begin{figure*}[!h]
\begin{center}
\includegraphics[scale=.48]{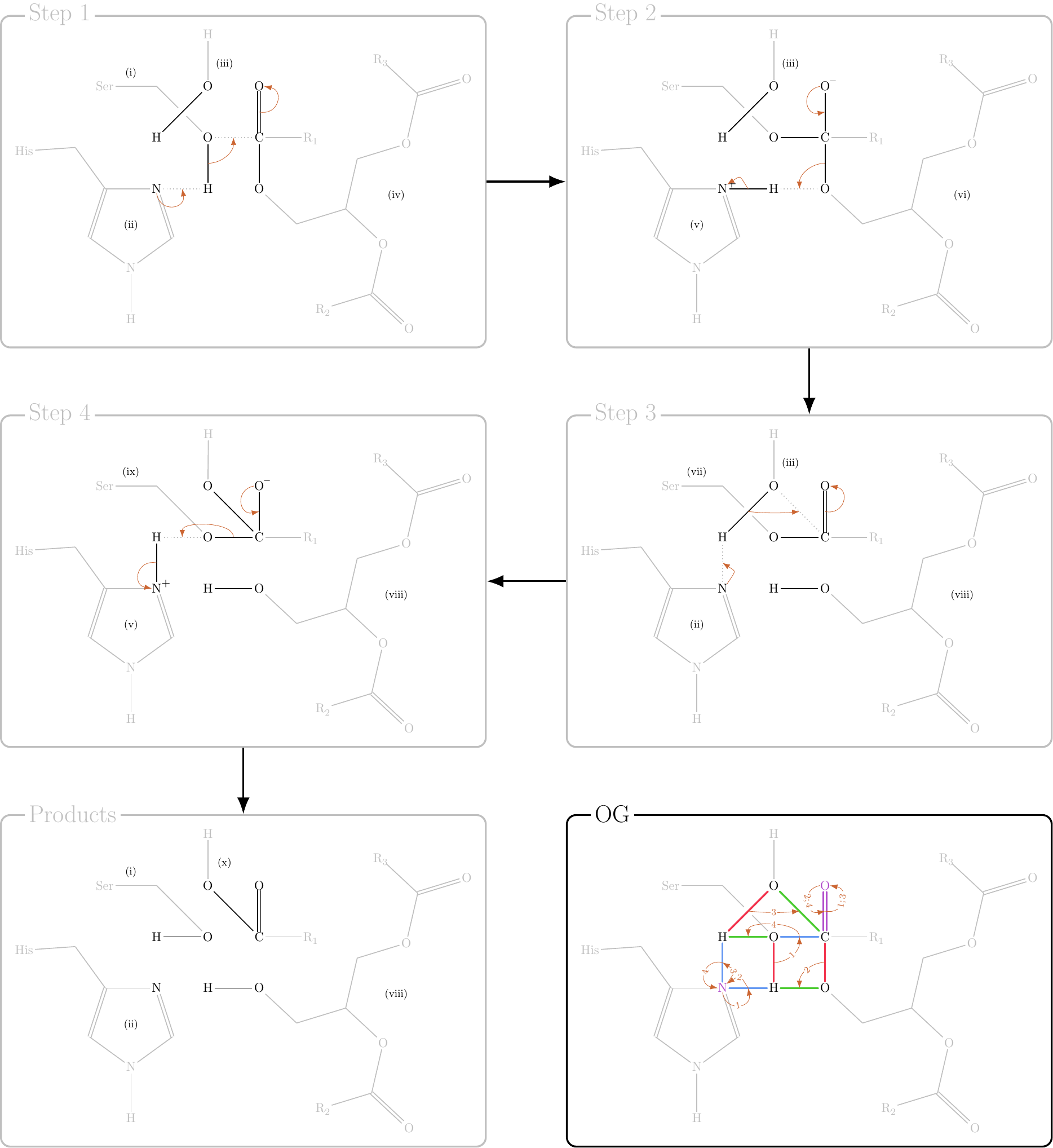}
\caption[Four-step mechanism]{The four-step mechanism (M-CSA \mcsaentry{218}) of pancreatic triacylglycerol lipase (\uniprot{P29183}). This \ECnumsimple{3}{1}{1}{3}~reaction catalyzes the conversion of a triacylglycerol (iv) into a fatty acid ion (x) and diacylglycerol (viii). In the first step, a histidine (ii) deprotonates a serine (i), which now attacks a carboxyl carbon of the substrate (iv). As a result, the serine becomes covalently bound to the substrate (vi), and the histidine side chain is protonated (v). In the second step, the oxyanion collapses, releasing diacylglycerol (viii) and deprotonating the histidine (ii). In the third step, the histidine (ii) deprotonates a water molecule (iii), enabling the water to attack the acylated serine (vii). In step four, the oxyanion collapses once more, ejecting a fatty acid ion (x) and reconstituting the chemical state of the amino acid side chains. The lower right panel depicts all steps stacked on top of each other, corresponding to the overlay graph, as explained in the main text. Bonds created or broken during the reaction are shown in green and red, respectively. Transient bonds that are first created and then broken are shown in blue, whereas transient bonds first broken (or modified) and then restored are shown in purple. Likewise, atoms that undergo transient state changes (such as partial charge) are also shown in purple. Atoms and bonds in gray are not directly involved in the reaction and are shown for context.}
\label{fig:mech_example}
\end{center}
\end{figure*}
\afterpage\clearpage

\begin{figure*}[!h]
\begin{center}
\includegraphics[scale=.9]{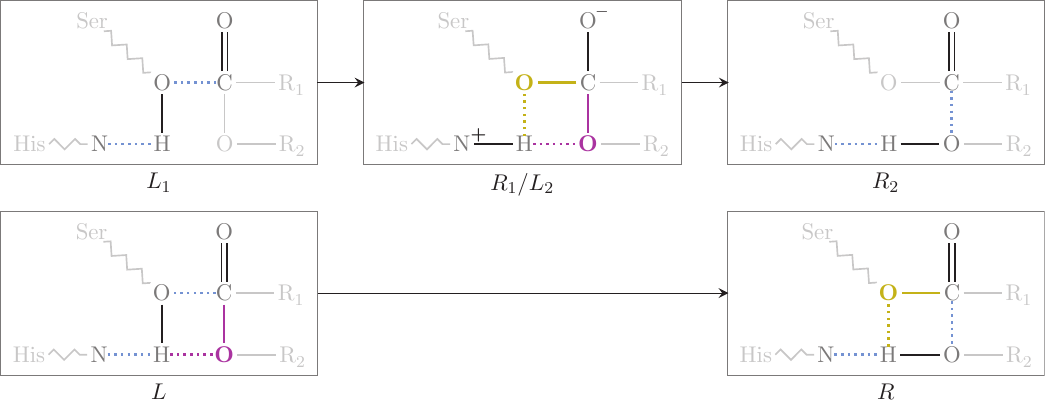}

\caption{The rules $p_1=L_1 \rightarrow R_1$ and $p_2=L_2\rightarrow R_2$ corresponding to steps 1 and 2 of the mechanism in~\hyperref[fig:mech_example]{Figure~\ref{fig:mech_example}}. The overlap of $R_1$ and $L_2$ is given by the relative position of the arrow pushes involved in the two steps. Yellow and magenta are used to highlight the sub-patterns unique to $p_1$ and $p_2$, respectively. The gray parts are not part of the rules and are only included to situate them in the mixture. The bottom row shows the rule obtained by the composition of $p_1$ and $p_2$. Conceptually, the left pattern $L$ of the composite rule $p$ must collect the conditions (the structural parts) required by the left pattern of $p_1$ plus those conditions of the left pattern of $p_2$ that have not been created by the action of $p_1$. Likewise, the right pattern $R$ of $p$ must contain the right pattern of $p_2$ plus those parts of the right pattern of $p_1$ that have not been modified by $p_2$. This is formalized in the \hyperref[app:rule_comp]{Supporting Information Section~\ref{app:rule_comp}}.}
\label{fig:composition}
\end{center}
\end{figure*}

\begin{figure*}[!ht]
\begin{center}
\includegraphics{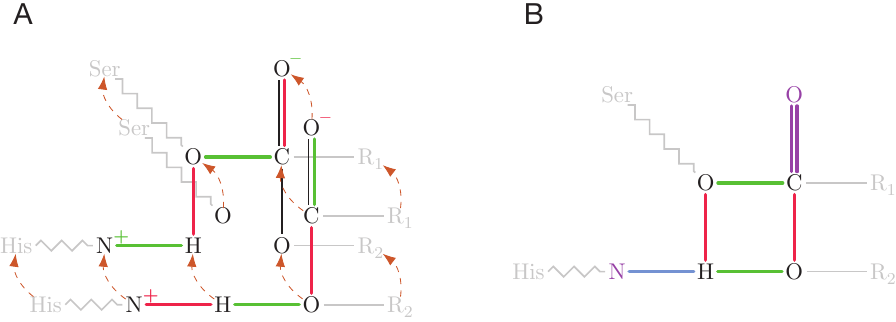}

\caption{{\bf A:} The graphical superposition of the ITS drawings associated with the rules derived from steps 1 and 2 of the mechanism in \hyperref[fig:mech_example]{Figure~\ref{fig:mech_example}}. The color coding is explained in the main text. The gray parts do not play a part in the superposition and are only used to localize the ITS elements within the reactants. {\bf B:} The resulting \ourITSname{} obtained by superimposing steps 1 and 2.}
\label{fig:overlay}
\end{center}
\end{figure*}

Our objective here is to generalize the ITS idea from one elementary reaction to a sequence of elementary reactions. As an example, consider the enzymatic multi-step mechanism depicted in~\hyperref[fig:mech_example]{Figure~\ref{fig:mech_example}}. (Here the \ch{-R1},\ch{-R2}, and \ch{-R3} symbols stand for actual structures that we abbreviated for the sake of less clutter.) Each step is represented as an arrow-pushing diagram and can be translated into a rule. The first two steps are shown in the top row of~\hyperref[fig:composition]{Figure~\ref{fig:composition}}, where the middle pattern depicts the overlap between the right side of the rule for step 1 and the left side of the rule for step 2. The bottom row shows the corresponding composite rule. The intuition behind rule composition is perhaps best illustrated with the help of the ITS drawings associated with each elementary step, as shown in~\hyperref[fig:overlay]{Figure~\ref{fig:overlay}A}. The ITS of step 2 is overlaid (red dashed arrows) onto that of step 1. The overlay makes the ITS colors \enquote{interact} when modifications done by rule 1 are undone by rule 2. Such modifications are transient. For example, a bond is broken only to be created again and charges that change in one step are restored in another. This is recorded by the color interaction. If only one ITS element is colored (no interaction) the color is simply inherited by the overlay: a bond or atom is shown in green when the net change is a creation and in red when the net change is an elimination. If red overlays on green, the color changes to blue, indicating bonds that are created and then eliminated again. If green overlays on red, the color turns purple, indicating bonds and atoms whose original state is altered and then restored. \enquote{State} here refers to bond type (single, double, triple) or charges. We refer to multiple ITS overlays as an overlay graph or \ourITSname. The \ourITSname{} corresponds to rule composition.

This process generalizes to $n$ steps. While fairly intuitive, the proper definition and automation of rule composition require a formalization, which we spell out in detail in the \hyperref[app:rule_comp]{Supporting Information (Section~\ref{app:rule_comp})}. The treatment presented there is meant to be a formal specification to guide implementation. At the time of this writing, our graph-transformation platform for chemistry~\cite{AndersenFMS2016} does not handle the non-bond constraints (dotted lines) in \emph{full} generality. The procedure with which we obtained the results in this paper is therefore a custom-implementation (script available online~\cite{implementation}) that makes use of the extant platform but does not extend it. A fully general treatment of chemical graph transformation as proposed in~\cite{BehrKAM2021} is a task for the future.

The \ourITSname{} (or, equivalently, the composite rule) underscores the mechanistic significance of molecular parts that appear unchanged at the resolution of an overall reaction. The utility of a composite rule becomes apparent upon considering that it can be applied to a mixture other than the original reactants. A match of the $L$-pattern of the composite rule guarantees that the chemical transformation of the new reactants is compatible with the mechanism represented by the rule. Composite rules, therefore, enable the search for potential reactants other than those from which they were abstracted. In particular, they can suggest the assignment of a mechanism to a reaction for which none is known. This differs from a previous approach~\cite{AndersenFFFKLMN2021} that searched for a combination of elementary steps to achieve the desired outcome. While the previous approach can propose novel mechanisms by recombining steps, it suffers from the usual combinatorial explosion, confining it to rather short mechanisms. Here, the composite rule is a unit whose underlying steps can no longer be shuffled, thus curtailing the combinatorial explosion in searching for compatible reactant/product pairs, but also limiting the present approach to known mechanisms. We illustrate applications in \hyperref[sec:results]{Section~\ref{sec:results}}.

A few remarks are in order to keep perspective. Our framework represents a microscopic view in which atoms have unique identifiers, which is distinct from a more coarse-grained view based on chemical equivalence. This affects the meaning of \enquote{change} or \enquote{modification}. For example, a bond remains (overall) unchanged only if it is broken and reconstituted between the \emph{same} partners at the microscopic level. This is useful in the context of enzyme reactions because it detects atoms exchanged between the enzyme and its substrates, although the enzyme remains chemically the same.

\begin{figure*}[!h]
\begin{center}
\includegraphics[scale=.8]{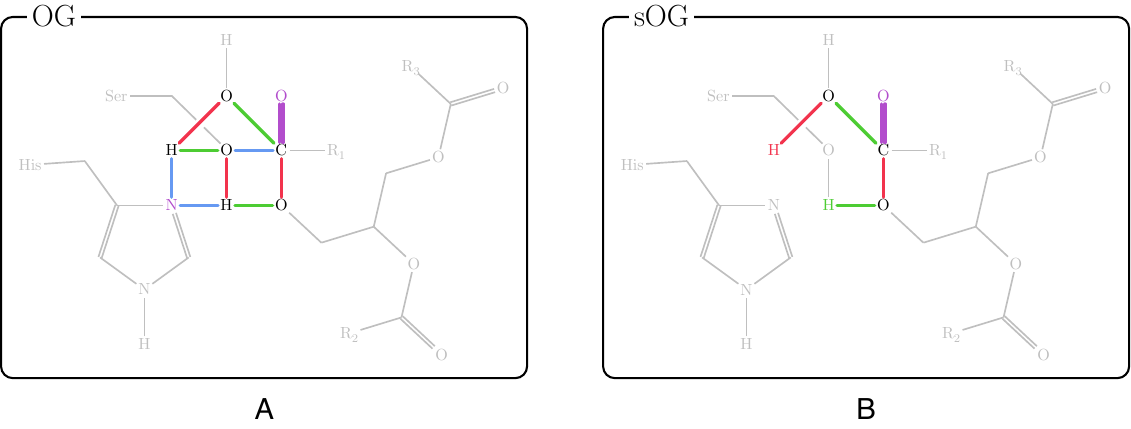}
\caption[Overlay graph]{{\bf A:} \ourITSname{} of the triglyceride lipase mechanism from \hyperref[fig:mech_example]{Figure~\ref{fig:mech_example}}. The color coding is detailed in the text. The gray parts of the molecules are only shown for context and are not components of the \ourITSname{} or the associated composite rule. In the present example, the catalyst is present only through a nitrogen atom of the histidine and the \ch{OH} group of the serine. {\bf B:} The substrate \ourITSname{} (s\ourITSname{}) derived by removing all atoms belonging to the catalyst. See text for details.}
\label{fig:its_and_substrate}
\end{center}
\end{figure*}

The \ourITSname{} is guaranteed to capture all atoms and bonds transiently involved in the reaction mechanism. Yet, the present formalism does not retain the exact nature of the transient modification in all cases. For instance, in~\hyperref[fig:its_and_substrate]{Figure~\ref{fig:its_and_substrate}A}, the double bond is flagged as transiently modified, but the more specific information that it has been reduced to a single bond twice is lost. Similarly, the \ch{N} of the histidine transiently changed charge, but the number of times it changed and the direction of state change (whether charge increased or decreased) are not recorded. In the final analysis, from a composite rule, as formalized here, we cannot recover the following information: (i) the history of temporary changes a bond underwent before restoration of the original state; (ii) the history of transient changes a bond underwent before a non-transient change; (iii) the transient type (single, double, triple) of a bond formed and eliminated again (blue bond in the \ourITSname). We detail these limitations in the \hyperref[app:limits]{Supporting Information Section~\ref{app:limits}}.

\subsection{Substrate Rules}
\label{sec:substrate_its}

In the case of catalysis, we can make a useful generalization by abstracting the identity of the catalyst in the composite rule to only keep those molecular parts of the substrate that are required in the interaction with the catalyst, given the mechanism captured by the rule. In other words, we can \enquote{anonymize} the catalyst without compromising the mechanism. By way of metaphor, consider a clip of a dance movement that requires two parties, say a waltz, but make one of the partners transparent to show only the movements of the other. In such a \enquote{minus one} clip, any other partner making waltz movements can be dropped in place of the transparent one.

More specifically, following~\hyperref[fig:mech_example]{Figure~\ref{fig:mech_example}}, consider the hydrogen atom that belongs to the serine of the catalyst. This \ch{H} is given to the substrate and a different \ch{H}, originally belonging to a water molecule and transiently parked at the \ch{N} of a histidine, is returned to the serine, restoring its chemical state. Clearly, this \ch{H} is essential for the reaction, but it does not strike one as essential that it must come from an \ch{O}. Any proton donor would do. A temporary parking space for hydrogens is essential for this mechanism, but, again, it seems less important that the parking lot be a \ch{N}. The broader context---that the \ch{O} belongs specifically to a serine and the \ch{N} to a histidine---is not represented anyway in these rules, at least not at the present level of abstraction. Likewise, the proton donor and the temporary parking lot need the right pK$_{\text{a}}$ to engage in catalysis, but this, too, is not within the scope of the simple graphical model used here.

These observations suggest a reduced (thus more general) composite rule in which only those parts of the substrates are retained that are required for their transformation. The enzyme and any other catalytic molecules are abstracted away. We refer to this reduced rule as a substrate rule or, in the overlay graph representation, as the substrate \ourITSname{} (s\ourITSname{} for short). The substrate rule has the advantage of being applicable to a mixture consisting only of substrate molecules, leaving the exact nature of the catalyst(s) unspecified, yet retaining the structural requirements from the mechanism. This means that any alternative catalysts that can be plugged into the reaction (that is, match the substrate rule) are potential catalysts for the transformation of the same substrate using the same mechanism.

\begin{figure*}[!h]
\begin{center}
\includegraphics[scale=.8]{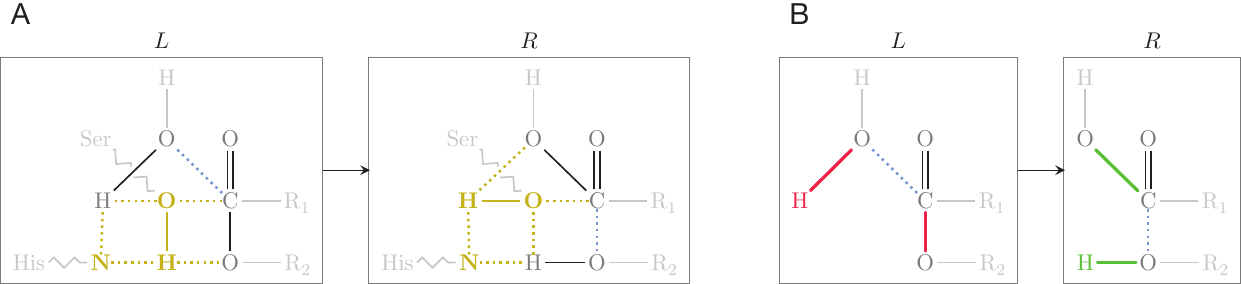}

\caption{{\bf A:} The composite rule of the full triglyceride lipase mechanism from \hyperref[fig:mech_example]{Figure~\ref{fig:mech_example}}. The catalytic components of the composite rule are highlighted in yellow. Notice that the atoms are not required to preserve their catalytic identity across the reaction. {\bf B:} The corresponding substrate rule, obtained by removing the highlighted catalytic components. Red and green are used for parts that are removed, respectively created, by the rule. The atoms that change their catalytic identity appear as being destroyed or created in the substrate rule.}
\label{fig:sub_rule}
\end{center}
\end{figure*}

To construct the substrate rule associated with a catalytic mechanism, we need to remove from the composite rule those parts representing the catalyst. This is illustrated in~\hyperref[fig:sub_rule]{Figure~\ref{fig:sub_rule}A} for the whole mechanism shown in~\hyperref[fig:mech_example]{Figure~\ref{fig:mech_example}}. (This composite rule differs from the one shown in~\hyperref[fig:composition]{Figure~\ref{fig:composition}}, which resulted from composing only step 1 and step 2.)

We follow the usual notion of a catalyst as a molecule that appears both on the educt and product side. The molecular parts of the catalyst have to be identified separately on the educt and the product side. These parts are highlighted in gold in~\hyperref[fig:sub_rule]{Figure~\ref{fig:sub_rule}A}. For instance, the hydrogen atom that is part of the serine before the reaction is different (in terms of microscopic identity) from the one attached afterwards. The substrate rule obtained by removing the highlighted atoms and bonds from the composite rule is shown in~\hyperref[fig:sub_rule]{Figure~\ref{fig:sub_rule}B}.

As a consequence of this resection, the left and right patterns of a substrate rule no longer have to contain the (microscopically) same atoms. Yet, mass conservation remains intact because $L$ and $R$ of the substrate rule contain the same number and type of atoms. For example, the substrate rule of~\hyperref[fig:sub_rule]{Figure~\ref{fig:sub_rule}B} means that the \ch{H} on the left, colored red, literally vanishes from the substrate system and the \ch{H} on the right, colored green, appears out of nothing. This is a consequence of having removed the catalyst from the picture. In a real-world setting, the \ch{H} that disappears must be received by a catalyst, while the \ch{H} that seems to be created out of nothing must be provided by the same catalyst. The point of the substrate rule is to express that the \ch{H} donor and acceptor can belong to any catalyst, not just the particular one whose mechanism gave rise to the composite rule. Moreover, if the \ch{H}-donor and \ch{H}-acceptor were not catalysts, then some additional molecules would have been modified and would have to show up in the rule. The substrate rule thus guarantees the catalytic nature of the transformation without specifying the catalyst.

We described the \enquote{typical} case and for the sake of readability glossed over some fine print on which we elaborate in the Supporting Information, Sections~\ref{app:substrate_rule} and \ref{app:catalyst_ambiguity}. Some of that fine print is purely technical, but some is conceptually interesting and worth a quick brush stroke. The formalization of the substrate rule tugs at the question of what, exactly, qualifies as a catalyst. Consider a situation in which an enzymatic process generates an intermediate product \ch{X} that then serves in a catalytic capacity for the remainder of the process by, say, helping restore the state of the enzyme \ch{E}. The overall reaction might read like \ch{E + S -> E + P + X}. Clearly, \ch{X} is not a catalyst by IUPAC standards, since it does not show up on the educt side. Yet, in the context of the mechanism, \ch{X} participates as a catalyst generated \emph{in situ}. The substrate rule formalism is not guaranteed to work in such cases without additional assumptions. An example, where \ch{X} is \ch{H2O}, is given in \hyperref[app:catalyst_ambiguity]{Section~\ref{app:catalyst_ambiguity} of the Supporting Information}. (With a bit of squinting, \hyperref[app:catalyst_ambiguity]{Section~\ref{app:catalyst_ambiguity}} can be read without plowing through the formalism.)

\section{Results and Discussion}
\label{sec:results}

\subsection{Obtaining Overlay Graphs}
\label{sec:building_its}

The M-CSA database~\cite{RibeiroHFTFT2017} is a non-redundant catalog of enzymatic reaction mechanisms represented as sequences of elementary steps annotated with electronic displacements in the style of arrow pushing. This kind of information is suited for constructing \ourITSname{}s, as in \hyperref[fig:mech_example]{Figure~\ref{fig:mech_example}}.

One problem is that reaction steps listed in the M-CSA do not explicitly specify atom maps. This does not present an issue for each step considered in isolation, since the flow of electronic displacements suffices to relate atoms in the educts to those in the products. Rather, the problem is the absence of atom maps across steps. While the visual display of an M-CSA mechanism might suffice for a human user to infer the connections between steps, visualization comes with connotations that are not always present in the underlying data and therefore not available to automated analysis.

\begin{figure}[!ht]
\begin{center}
\includegraphics[scale=.8]{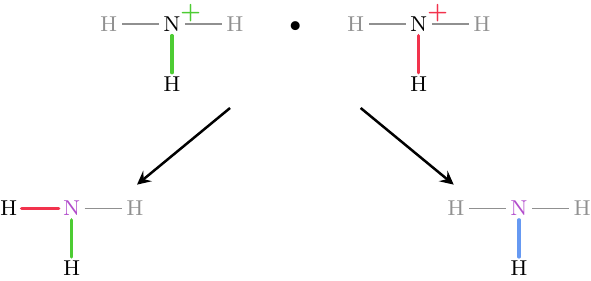}
\caption{The left pattern of a rule removing a hydrogen atom can be compatible with any of the three hydrogen atoms of the \ch{H3N+} group. Depending on the choice of match, two distinct rule compositions are possible, resulting in different \ourITSname{}s.}
\label{fig:symmetry}
\end{center}
\end{figure}

The absence of a specific atom map connecting the steps of an M-CSA mechanism opens the possibility for the left pattern of one step to match the products of the previous step in several ways, should these products contain multiple equivalent instances of a chemical substructure, such as the hydrogen atoms of a methyl or an amino group. Likewise, the presence of several instances of the same molecular species, like \ch{H2O}, during a mechanism yields a choice of which instance to use in a given step. Such symmetries are relevant for the construction of the composite rule only if they involve substructures with distinct histories. For example, \hyperref[fig:symmetry]{Figure~\ref{fig:symmetry}} illustrates how the atom map linking two steps impacts the composite rule as represented by its \ourITSname{}. The first step attaches a hydrogen to an amine group, \ch{H2N}, while the second cleaves a hydrogen atom from the resulting \ch{H3N+} group. Since all three hydrogen atoms of that group are equivalent, the left pattern of the second rule can match any one of them. However, since the hydrogen atoms that came with the initial \ch{H2N} have necessarily acquired a different history from the hydrogen atom that was just added to form the \ch{H3N+}, we obtain two non-isomorphic mechanisms and associated \ourITSname{}s, depending on which class the cleaved hydrogen belongs to.

Rule composition is based on an underlying sequence of rule applications (the mechanism), each of which involves a particular choice of match into the mixture. A specific sequence of matches, along with the atom identities across each step, defines the atom map for the entire mechanism. To avoid confusion, we refer to the series of steps in an M-CSA entry as a \enquote{sequence} and reserve the term \enquote{mechanism} for a sequence with an atom map. Since for many M-CSA entries several matches are possible at each step, we consider all of them and treat them as distinct mechanisms.

At the time of writing, the M-CSA database contains \num{888} sequences of steps~\cite{mcsa_api}. We were able to generate \num{2802} mechanisms, and attendant composite rules and \ourITSname{}s, for a total of \num{600} M-CSA sequences. \num{288} M-CSA sequences could not be cast as rules due in part to current limitations of the implementation of chemical graph transformation, which, for example, does not yet process radical formation, and in part due to difficulties with extracting the needed information from some M-CSA entries. For the process used to prepare the M-CSA information for our approach, see~\cite{AndersenFFFKLMN2021}. Of these \num{600} M-CSA sequences, \num{435} have a unique mechanism and \num{100} have two mechanisms. Five M-CSA sequences are compatible with over \num{100} different mechanisms. One sequence, M-CSA \mcsaentry{38}, gives rise to \num{964} non-isomorphic composite rules because it involves four water molecules that can become distinguishable as the mechanism proceeds.

The number of possible non-isomorphic rule compositions grows exponentially with the number of disjoint isomorphic substructures. Thus, even a partial atom map that breaks some symmetries can reduce the number of \ourITSname{}s substantially. For example, M-CSA \mcsaentry{186} has \num{24} possible mechanisms that can be reduced to a single \ourITSname{} by mapping just four atoms, each in one intermediate.

\subsection{Interpretation of \ourITSname{}s}
\label{sec:its_graph_interpretation}

The construction of a composite rule, and by implication its \ourITSname{}, preserves information about transient changes. If a mechanism includes some steps that are independent of others, their relative order of temporal execution is not fixed. Naturally, the compression of a sequence of steps into a single overall rule loses information about the ordering of such steps, but by their independence, this ordering is irrelevant for the outcome. Independence is often the case for \enquote{clean-up} steps that reconstitute the chemical state of the enzyme. At the end of Section 2.1 we summarized the kind of information that is lost by the composite rule or the \ourITSname{}.

To determine whether the information preserved in an \ourITSname{} provides insights into enzymatic reaction mechanisms, we constructed the \ourITSname{}s of reactions within the \ECnum{3}{1}{1}{} subclass of which the mechanism depicted in \hyperref[fig:mech_example]{Figure~\ref{fig:mech_example}} is an example. The first position in the EC numbering scheme (i.e.~\ECnum{3}{}{}{}) refers to the hydrolytic cleavage of bonds. This process transfers a substrate group to specifically \ch{H2O} as the acceptor. The second position (\ECnum{3}{1}{}{}) indicates the nature of the hydrolysed bond; in the present case, an ester bond. The third position (\ECnum{3}{1}{1}{}) specifies the nature of the substrate, here a carboxylic ester. The M-CSA lists \num{24} reaction sequences as \ECnum{3}{1}{1}{}. Converting these sequences into mechanisms does not reveal much ambiguity: Only two reaction sequences allow for more than one \ourITSname{}---specifically, \num{2} and \num{5} \ourITSname{}s---yielding a total of \num{29} \ourITSname{}s. The ambiguities arise from the equivalence of hydrogen atoms in an \ch{H2O^{+}} group and an \ch{H3N^{+}} group, similar to the case illustrated in \hyperref[fig:symmetry]{Figure~\ref{fig:symmetry}}.

\begin{figure*}[!h]
\begin{center}
\includegraphics[scale=.8]{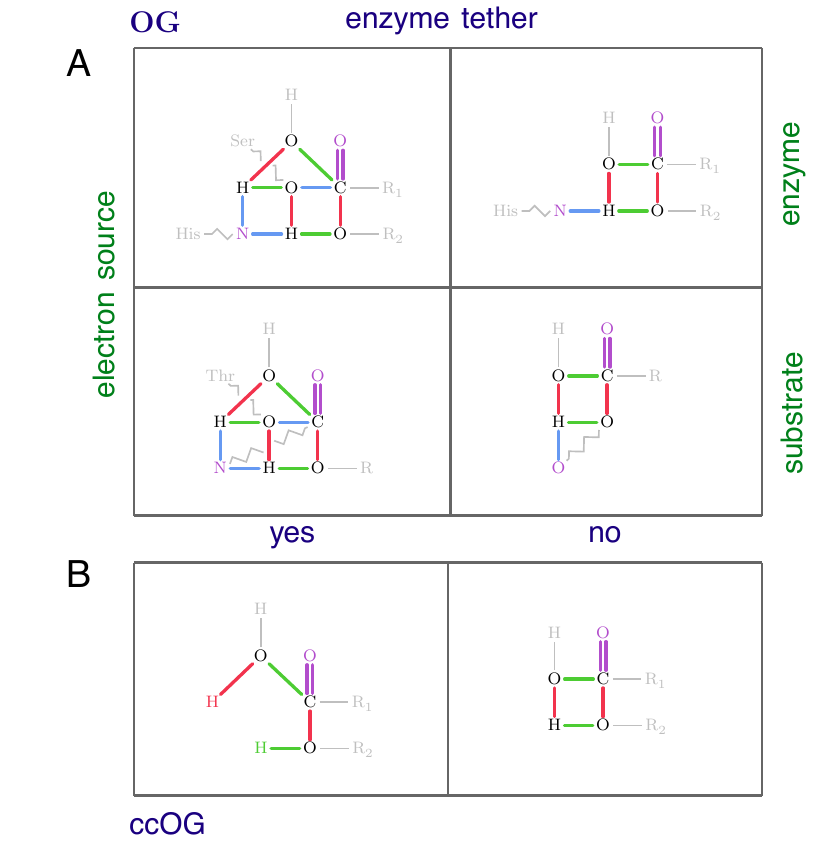}
\caption[Overlay graph patterns]{{\bf A:} Classification of different \ourITSname{} patterns extracted from \ECnumsimple{3}{1}{1}{-} reactions listed in the M-CSA.  {\bf B:} Catalytic complement overlay graphs, derived from \ourITSname{}s by removing catalytic sites regardless of whether they belong to a distinct catalyst molecule or the substrate or both. These cc\ourITSname{}s distinguish between the equivalence classes along the tether/no-tether axis.}
\label{fig:311_comp}
\end{center}
\end{figure*}

The set of \ECnum{3}{1}{1}{} reactions indeed translates into patterns at the level of \ourITSname{}s (\hyperref[fig:311_comp]{Figure~\ref{fig:311_comp}}) revealing two mechanistic distinctions that organize enzyme-substrate relations into four classes. 

The first distinction refers to how many catalytic sites engage in a chemical interaction with the substrate. In the context of \ECnum{3}{1}{1}{} reactions, this amounts to whether two or one hydrogen atoms are involved in the catalytic process, which in turn indicates whether or not the mechanism proceeds via a covalently bound intermediate. In the case of a bound intermediate, the flow of electronic displacements repeats twice; once to create the covalent bond to the enzyme followed by the release of one product, and a second time to break that covalent bond, replacing it with a bond to the \ch{OH} of water, thereby releasing the second product of hydrolysis. Each flow is a round trip of electrons from a source to a sink and back, each round trip involving a distinct proton that engages in a transient bond. Absent a covalently bound intermediate only one round trip is necessary, requiring only one proton. 

The second distinction pertains to whether the source of electrons that initiates the reaction sequence is located on the enzyme or the substrate. At the abstraction level of arrow pushing, an elementary step is a concerted electronic displacement prompted by a region of high electron density sufficiently \enquote{motivated} by a chemical environment. Thus, a special role falls to any step in a catalytic process that initiates an electronic displacement sequence from a relatively stable intermediate, including, in particular, the very first step of the reaction. 

\hyperref[fig:311_comp]{Figure~\ref{fig:311_comp}} arrays the four \ourITSname{} patterns according to electron source and whether tethering of an intermediate occurs. The upper left quadrant is our running example of \hyperref[fig:mech_example]{Figure~\ref{fig:mech_example}}, \mcsaentry{218}. The reaction mechanism generates a covalently bound intermediate with the enzyme as the initial electron source. This pattern is present in \num{19} of the \num{29} \ECnum{3}{1}{1}{} \ourITSname{}s. These \num{19} \ourITSname{} are derived from as many M-CSA reaction sequences with no ambiguity in the underlying mechanisms. The lower left quadrant shows the case of a D-aminoacyl-tRNA deacylase, \mcsaentry{748} (sequence proposal \#1), in which the tethering occurs to a threonine of the enzyme, but the electron source is provided by the \ch{N} of an amino group bound to a \ch{C} in $\alpha$ position to the carbonyl group of the substrate. The upper right quadrant shows \mcsaentry{83}, phospholipase A2, whose hydrolase action proceeds without creating a covalently tethered intermediate and uses a histidine residue as initial electron source. Although no covalent tethering occurs, the substrate is nonetheless locked in position by coordination to \ch{Ca^2+}. At the abstraction level of \ourITSname{}s, such bonds are not considered as properly catalytic with regard to the actual chemical transformation, but rather as screening and/or securing a substrate in the catalytic pocket. These interactions are presumably critical for catalytic efficiency and thus belong to the catalytic process in a broader sense that could, in principle, be captured by rules and attendant \ourITSname{}s with substantially more context beyond the reaction center. Yet, such extensive and specific rule refinements would blur the general picture, making \ourITSname{}s less useful. Finally, the lower right quadrant is again the case of the D-aminoacyl-tRNA deacylase, \mcsaentry{748}, but this time based on an alternative reaction sequence proposal listed in the M-CSA. In this case, the hydrolysis is entirely intra-molecular to the substrate, with the enzyme just providing a tailored physico-chemical environment. This particular reaction sequence gives rise to two mechanisms, and thus \ourITSname{}s, due to the equivalent hydrogen atoms in the protonated \ch{OH}-group after the first step. It should be noted, however, that the M-CSA lists the reaction sequences for \mcsaentry{748} with a low confidence rating.

Especially in the context of enzymatic reactions, the notion of a catalyst typically connotes an entity distinct from the substrate. The above observations, however, show that catalytic functionality can be distributed across molecular entities and may, therefore, cross-cut the traditional separation of enzyme and substrate, even though known cases seem rare so far. If we drop the notion of a unitary catalyst in favor of a collection of \emph{catalytic sites} that could be distributed across the theater of a chemical reaction, then our construction of removing atoms carrying catalytic functionality from the \ourITSname{} to produce a substrate \ourITSname{} would, in the case of the lower quadrants in \hyperref[fig:311_comp]{Figure~\ref{fig:311_comp}A}, remove atoms belonging to the substrate itself. This generalization suggests calling the resulting construct a \enquote{catalytic complement} overlay graph or cc\ourITSname{} (\hyperref[fig:311_comp]{Figure~\ref{fig:311_comp}B}) rather than a \emph{substrate} \ourITSname{}. The cc\ourITSname{} of the upper left quadrant is the same as the s\ourITSname{}, since the substrate carries no catalytic sites. Yet, the cc\ourITSname{} of the lower left quadrant is isomorphic to that of the upper left. Similar observations hold for the right two quadrants. The cc\ourITSname{} appears to classify mechanisms according to the number of catalytic sites involved.

It is tempting to speculate that the evolution of some enzymes might involve shifting catalytic functionality from substrate to enzyme or vice versa. For example, the efficiency and reliability of an originally self-catalyzed intra-molecular reaction (such as the lower right quadrant) might be improved by moving sites of catalysis from the substrate to a polypeptide pocket that originally provided no more than a suitable environment for the reaction.

\subsection{Using substrate rules to Identify Compatible Reactions}
\label{sec:matching_its_rules}

The applicability of a composite rule to a mixture guarantees that the chemical mechanism represented by the rule is feasible in the mixture. Feasibility must be understood at the abstraction level defined by graph transformation, which, as it stands, has nothing to say about thermodynamics. With this in mind, we can use a composite rule to screen for reactions whose educts and associated products match the rule's left and right patterns, respectively, to suggest a mechanism, when none is known, or propose an alternative to a known mechanism. To be clear, a composite rule is applicable much like any other rule to a variety of reactants matching its left pattern. The point here is that the composite rule represents a series of steps and can be constructed programmatically, as shown in earlier sections, thus automating the search for educt-product pairs that are compatible with a given mechanism.

To this end we extracted over \num{13000} overall reaction descriptions from the Rhea database, an expert-curated database of biologically interesting reactions~\cite{LombardotMAAHNIXCRB2019}. Since these descriptions do not explicitly list catalytic components, we deploy substrate rules, which capture the mechanism while minimizing the specification of the catalyst. To test the approach we use substrate rules derived from the composite rules that represent the instances in the \ECnum{3}{1}{1}{} reaction class of the M-CSA. The \num{29} composite rules and associated \ourITSname{}s characterized in the previous section reduce to \num{11} equivalence classes at the level of substrate rules and s\ourITSname{}s: one class of size \num{18}, one of size \num{2}, and nine of size \num{1}. Of these \num{11} distinct substrate rules, \num{3} are refinements of others; that is to say: in \num{3} cases, when one rule applies, another one, strictly less selective, applies as well and produces the same result. We therefore discard the more selective rules. One substrate rule came from para-nitrobenzyl esterase, which cleaves a \ch{C-N} bond, making its \ECnum{3}{1}{1}{} classification uncertain. We opted to leave it out, ending up with \num{7} distinct substrate rules capturing \ECnum{3}{1}{1}{} mechanisms.

The M-CSA and Rhea databases use different conventions for carboxylic acids. For each reaction marked as \ECnum{3}{1}{1}{}, the M-CSA writes a carboxylic acid group as \ch{COOH}, whereas Rhea writes it as \ch{COO-} plus \ch{H+}. Since our rules are derived from the M-CSA, we need to additionally cleave the \ch{O-H} bond of the carboxylic acid in the product. Recall that this \ch{OH} group comes from a water molecule on the educt side. In rules that are minimal in the sense of only representing the reaction center, the water molecule need not be fully specified. For example, the rule that captures the mechanism of M-CSA \mcsaentry{218} and is shown in \hyperref[fig:sub_rule]{Figure~\ref{fig:sub_rule}A} specifies one \ch{H-O} bond of water in its left pattern, but not the other \ch{O-H} bond even though the latter ends up in a product molecule. The reason is that nothing happens to that other \ch{O-H} bond; it simply gets transferred along as a result of the creation of a bond between the carbon of the carboxyl group and the \ch{O} of the water. However, if that \enquote{inherited} bond is cleaved in a subsequent step, in order to achieve the carboxylate form, the modified rule now must fully specify the water molecule. Since a larger left pattern makes rules more specific, we also keep the \num{7} original unmodified rules.

These \num{14} substrate rules match \num{588} Rhea reactions, which are, therefore, compatible with the catalytic mechanism encapsulated in the matching rules. We utilize these matches for two types of analyses. To evaluate the retrieve rate of the known \ECnum{3}{1}{1}{} reactions and to explore reactions of different or unknown EC class which are compatible with an \ECnum{3}{1}{1}{} mechanism.

A total of \num{130} reactions are marked as \ECnum{3}{1}{1}{} within the Rhea database. Out of these, we retrieve \num{90}, giving us a recall rate of \SI{69.2}{\percent}. All these matches are, as expected, against the modified rules. The reasons for not matching the remaining \num{40} reactions are manifold. \num{20} reactions exhibit an oligomeric nature not reflected in the SMILES representation from which we constructed the molecule graphs, or they show educts and products using mismatched abstractions or tautomerisations; \num{5} reactions involve interactions with iron, which are not supported by our present version of the graph-transformation approach to chemistry. The remaining \num{15} reactions do not stop after cleavage of the carboxylic ester bond but exhibit \enquote{follow-up} reactions of the kind listed in \hyperref[tab:unmatched_rhea]{Table~\ref{tab:unmatched_rhea}}.

\begin{table}
	\centering

	\small

	\begin{tabular}{@{}lr@{}}
		\toprule
		Description & Incidence \\
		\midrule
		Keto-enol tautomerionesation & \num{5} \\
		Second carboxylic ester bond cleavage  & \num{3} \\
		Decarboxylation  & \num{3} \\
		Cyclisation & \num{2} \\
		Intramolecular proton transfer & \num{1} \\
		Release of sulfite group & \num{1} \\
		\bottomrule
	\end{tabular}
	\caption{Types of additional modifications to the product molecules of the carboxylic ester bond cleavage observed in the Rhea database, with incidence.}
	\label{tab:unmatched_rhea}
\end{table}

Among these follow-up reactions, we sought to include the keto-enol tautomerisation as the most prevalent. The M-CSA database contains two elementary reaction steps corresponding to keto-enol tauromerisation, step \num{4} in \mcsaentry{85} and step \num{4} in \mcsaentry{463}, which are both described as spontaneous tautomerisation occurring outside the active site of the enzyme. We therefore augmented our rules by composing them with the steps corresponding to spontaneous keto-enol tautomerisation as present in the M-CSA. These steps were assumed to involve the ester oxygen, which ends up as a ketone group instead of an alcohol group. The augmented rules now match four of the five Rhea reactions in this category. In the remaining reaction, the keto-enol tautomerisation occurs twice; one involves the former ester oxygen and the other involves an alcohol group already present in an educt molecule. Other than that, the augmented rules do not match more Rhea reactions.

From among the remaining \num{498} matched reaction, not marked as \ECnum{3}{1}{1}{}, \num{30} are given a different EC number in the Rhea database and \num{468} reactions are assigned no EC number in Rhea. These results are summarized in \hyperref[tab:rhea_ec]{Table~\ref{tab:rhea_ec}}.

\begin{table*}
	\centering

	\begin{tabular}{@{}lrl@{}}
		\toprule
		EC class           & Incidence & Description                                     \\
		\midrule
		None               & \num{468} &                                                 \\
		\ECnum{3}{1}{1}{}  &  \num{90} & Carboxylic ester hydrolases                     \\
		\ECnum{2}{3}{1}{}  &  \num{19} & Acyltransferases other than amino-acyl          \\
		\ECnum{3}{6}{1}{}  &   \num{3} & Phosphorus-containing acid anhydrate hydrolases \\
		\ECnum{2}{3}{2}{}  &   \num{2} & Aminoacyltransferases                           \\
		\ECnum{4}{2}{99}{} &   \num{2} & Other carbon-oxygen lyases                      \\
		\ECnum{5}{4}{1}{}  &   \num{2} & Intramolecular acyl transferases                \\
		\ECnum{2}{1}{3}{}  &   \num{1} & Carboxy-/carbomyltransferases                   \\
		\ECnum{3}{2}{1}{}  &   \num{1} & Glycosidases                                    \\
		\bottomrule
	\end{tabular}
	\caption{The EC classification, as declared in the Rhea database, of the reactions explained by our substrate overlay rules.}
	\label{tab:rhea_ec}
\end{table*}

The majority of reactions classified as different class than \ECnum{3}{1}{1}{} are transferases, which cleave an ester bond and create a new one using an alcohol group of another educt molecule, as in \ch{R-CO-OR' + OH-R''-> R-CO-OR'' + OH-R'}. The alcohol group acts in place of the \ch{OH} of water in the hydrolase context, where \ch{R''=H}. These reactions resulted from a match to the unmodified substrate rules, since the water molecule is fully specified in the modified version.

Cross-linking the \num{468} reactions of unspecified EC class with UniProt~\cite{UniProt2020} reveals \num{310} reactions associated with at least one enzyme classified as \ECnum{3}{1}{1}{}. \num{200} of those are associated with exclusively \ECnum{3}{1}{1}{} enzymes, and \num{26} reactions are associated with no \ECnum{3}{1}{1}{} enzyme in UniProt. Similarly to the results in \hyperref[tab:rhea_ec]{Table~\ref{tab:rhea_ec}}, most of the retrieved enzymes outside of \ECnum{3}{1}{1}{} are acyltransferases (\ECnum{2}{3}{1}{}). Other prominent classes contain thiolester hydrolases (\ECnum{3}{1}{2}{}) and carbon-nitrogen hydrolases (\ECnum{3}{5}{1}{}). We retrieved no UniProt reference for the remaining \num{132} reactions.


\section{Summary and Conclusions}

Our starting point is the representation of chemistry within the framework of graph transformation, enabled by an open software platform developed for this purpose~\cite{AndersenFMS2016}. The widespread availability of versatile open-source tools for organic chemistry (e.g.~\cite{RDKit}) based on a linear string encoding of molecular structures begs the question of why a framework based on graph transformation is needed. An answer implicit in the present contribution is that the graph-transformation approach forces inclusion of the atom map. By virtue of the atom map, rules (of graph transformation) capture a mechanistic aspect of chemical reactions, albeit a very limited one. Nothing is said about the ordering of electronic displacements for complex mechanisms or about quantitative aspects that depend on geometry and thermodynamics. Mathematical graphs are not two-dimensional structures, they simply represent connectivity. While the framework can, in principle, be augmented to include information about quantitative aspects, the current approach has already demonstrated theoretical and practical utility~\cite{SilvaAHWBSF2019,AndersenFMS2015,AndersenMR2019,AndersenFFKMS2018,hcn:13,flow,AndersenFFFKLMN2021} to which the present work seeks to add.

We construct the composite rule of a sequence of rules derived from elementary steps---elementary in the sense of arrow pushing. The composite rule represents the \enquote{overall reaction}. A composite rule records not only the net changes between educts and products, but also transient changes, such as the temporary making or breaking of a bond or the temporary loss or acquisition of a formal charge. We visualize the information retained by a composite rule in terms of a single graph---the overlay graph---that superposes educt and product pattern along with a color-coded \enquote{fate map} of bonds reconstructed from the composite atom-map. The overlay graph can be viewed as the representation of a reaction mechanism using a graph-edit notation and constitutes a generalization to multi-step reactions of the imaginary transition structure proposed by Fujita for elementary reactions~\cite{Fujita1986}.

We pursue two specific applications in the context of enzymatically catalyzed reactions, enabled by a valuable resource of enzyme mechanisms known as the Mechanism and Catalytic Site Atlas (M-CSA). In the first case, we automatically construct the overlay graphs of all carboxylic ester hydrolase mechanisms (\ECnum{3}{1}{1}{}) listed in the M-CSA that had sufficient information to be translated into our framework. The compact nature of overlay graphs enables a visual inspection, which made us realize that these graphs can be classified according to whether the mechanism proceeds by tethering an intermediate product and whether the electronic displacement step initiating the reaction sequence resides within the substrate or the enzyme. This suggests that overlay graphs could be a useful and interpretable data type for processing complex reactions with statistical learning algorithms.

Any reaction whose educts and products match a composite rule can proceed by the mechanism the rule encapsulates. This provides a means for suggesting mechanisms of catalysis for reactions whose mechanism is unknown or for whom alternative mechanisms are sought. Since the composite rule includes catalytic parts that interact with the reaction center, candidate reactions listed in databases without specification of the catalyst would not be matched. To this end, we generalize the composite rule by eliminating all reference to the catalyst while retaining the \enquote{footprint} of catalytic action on the portion of the reaction center belonging to the substrate. With this construct, termed substrate rule (and the associated substrate overlay graph), we can identify reactions with unknown mechanism or unknown catalyst that could proceed with the captured mechanism. We exemplify this use case by screening all reactions in the Rhea database using the substrate rules derived from \ECnum{3}{1}{1}{} reactions in the M-CSA collection. We recover a substantial amount of reactions that were indeed classified as \ECnum{3}{1}{1}{} in Rhea, alongside many reactions that had no classification but whose study suggested compatibility with an \ECnum{3}{1}{1}{} mechanism. This illustrates the power of rule composition for automating the search for reactions that qualify as candidates for a given mechanism.

To end on a conceptual note, the idea behind the substrate rule is to capture the requirements of a specific catalyzed mechanism while being as agnostic as possible about the chemical implementation of the catalyst. This often works, but the graph transformation framework requires deciding who the catalyst is. A catalyst is usually viewed as a unitary object present on the educt side of a reaction and reconstituted on the product side. If all cases were like this, the construction of the substrate rule would pose no problem. A difficulty arises, however, in cases in which \enquote{the} catalyst is actually a collection of catalytic \emph{sites} that are distributed across the reaction theater, including the substrate itself. Even more vexing is the case of multi-step reactions where in addition to a catalyst present initially, a molecule (or functional group) is formed at an intermediate step of the reaction process and subsequently participates as a catalyst in completing the reaction. Such a \enquote{half-time} catalyst would show up as a product of the overall reaction but not as an educt and thus not qualify as a catalyst despite its crucial catalytic contribution.

\section*{Author Information}
\subsection*{Corresponding Authors}
{\renewcommand\labelitemi{}
	\begin{itemize}[leftmargin=*]
	    \item \acsauthor{Juri Kol\v{c}\'ak}{\denmark~and~\boston}{0000-0002-9407-9682}; Email: \href{mailto:juri.kolcak@gmail.com}{juri.kolcak@gmail.com}
		\item \acsauthor{Daniel Merkle}{\denmark}{0000-0001-7792-375X}; 
		Email: \href{mailto:daniel@imada.sdu.dk}{daniel@imada.sdu.dk}
	\end{itemize}
}
\subsection*{Authors}

{\renewcommand\labelitemi{}
	\begin{itemize}[leftmargin=*]
		\item \acsauthor{Jakob L. Andersen}{\denmark}{0000-0002-4165-3732}
		\item \acsauthor{Rolf Fagerberg}{\denmark}{0000-0003-1004-3314}
		\item \acsauthor{Christoph Flamm}{\vienna}{0000-0001-5500-2415}
		\item \acsauthor{Walter Fontana}{\boston}{0000-0003-4062-9957}
		\item \acsauthor{Christophe V.F.P. Laurent}{\denmark}{0000-0002-9112-6981}
		\item \acsauthor{Nikolai N{\o}jgaard}{\denmark}{0000-0002-7053-4716}
	\end{itemize}
}

\section*{Acknowledgments}
This work is supported by the Novo Nordisk Foundation grants NNF19OC0057834 and NNF21OC0066551 and by the Independent Research Fund Denmark, Natural Sciences, grants DFF-0135-00420B and DFF-7014-00041.

\bibliographystyle{ieeetr}
\bibliography{manuscript}

\begin{thebibliography}{10}

\bibitem{Fujita1986}
S.~Fujita, ``Description of organic reactions based on imaginary transition
  structures. 1. introduction of new concepts,'' {\em J. Chem. Inf. Comput.
  Sci.}, vol.~26, pp.~205--212, 11 1986.

\bibitem{RibeiroHFTFT2017}
A.~J.~M. Ribeiro, G.~L. Holliday, N.~Furnham, J.~D. Tyzack, K.~Ferris, and
  J.~M. Thornton, ``{Mechanism and Catalytic Site Atlas (M-CSA): a Database of
  Enzyme Reaction Mechanisms and Active Sites},'' {\em Nucleic Acids Research},
  vol.~46, no.~D1, pp.~D618--D623, 2017.

\bibitem{LombardotMAAHNIXCRB2019}
T.~Lombardot, A.~Morgat, K.~B. Axelsen, L.~Aimo, N.~Hyka-Nouspikel,
  A.~Niknejad, A.~Ignatchenko, I.~Xenarios, E.~Coudert, N.~Redaschi, and
  A.~Bridge, ``{Updates in Rhea: SPARQLing Biochemical Reaction Data},'' {\em
  Nucleic Acids Research}, vol.~47, no.~D1, pp.~D596--D600, 2019.

\bibitem{Levy2011}
D.~E. Levy, {\em Arrow-Pushing in Organic Chemistry}.
\newblock John Wiley \& Sons, 2011.

\bibitem{EhrigEGT2006}
H.~Ehrig, K.~Ehrig, U.~Golas, and G.~Taentzer, ``Fundamentals of algebraic
  graph transformation,'' in {\em Monographs in Theoretical Computer Science.
  An EATCS Series} (W.~Brauer, G.~Rozenberg, and A.~Salomaa, eds.), Berlin and
  Heidelberg, Germany: Springer, 2006.

\bibitem{HabelMP2001}
A.~Habel, J.~M{\"u}ller, and D.~Plump, ``Double-pushout graph transformation
  revisited,'' {\em Mathematical Structures in Computer Science}, vol.~11,
  no.~5, pp.~637--688, 2001.

\bibitem{AndersenFMS2016}
J.~L. Andersen, C.~Flamm, D.~Merkle, and P.~F. Stadler, ``{A Software Package
  for Chemically Inspired Graph Transformation},'' in {\em Graph
  Transformation. ICGT 2016. Lecture Notes in Computer science} (R.~Echahed and
  M.~Minas, eds.), vol.~9761, pp.~73--88, Cham, Switzerland: Springer, 2016.

\bibitem{bp2018}
N.~Behr and P.~Sobocinski, ``{Rule Algebras for Adhesive Categories},'' in {\em
  27th EACSL Annual Conference on Computer Science Logic (CSL 2018)} (D.~Ghica
  and A.~Jung, eds.), vol.~119 of {\em Leibniz International Proceedings in
  Informatics (LIPIcs)}, (Dagstuhl, Germany), pp.~11:1--11:21, Schloss
  Dagstuhl--Leibniz-Zentrum fuer Informatik, 2018.

\bibitem{AndersenFMS2018}
J.~L. Andersen, C.~Flamm, D.~Merkle, and P.~F. Stadler, ``Rule composition in
  graph transformation models of chemical reactions,'' {\em MATCH,
  Communications in Mathematical and in Computer Chemistry}, vol.~80, no.~3,
  pp.~661--704, 2018.

\bibitem{implementation}
J.~Kol\v{c}\'ak, C.~V. F.~P. Laurent, and N.~N{\o}jgaard, ``Overlay graph
  source code,'' 2022.

\bibitem{BehrKAM2021}
N.~Behr, J.~Krivine, J.~L. Andersen, and D.~Merkle, ``Rewriting theory for the
  life sciences: A unifying theory of ctmc semantics,'' {\em Theoretical
  Computer Science}, vol.~884, pp.~68--115, 2021.
\newblock TR: https://arxiv.org/abs/2106.02573.

\bibitem{AndersenFFFKLMN2021}
J.~L. Andersen, R.~Fagerberg, C.~Flamm, W.~Fontana, J.~Kolčák, C.~V. F.~P.
  Laurent, D.~Merkle, and N.~Nøjgaard, ``{Graph Transformation for Enzymatic
  Mechanisms},'' {\em Bioinformatics}, vol.~37, no.~Supplement 1,
  pp.~i392--i400, 2021.

\bibitem{mcsa_api}
EMBL-EBI, ``M-csa api,'' Accessed on Sep 24, 2021.

\bibitem{UniProt2020}
{The UniProt Consortium}, ``{UniProt: the Universal Protein Knowledgebase in
  2021},'' {\em Nucleic Acids Research}, vol.~49, pp.~D480--D489, 11 2020.

\bibitem{RDKit}
G.~Landrum, ``Open-source cheminformatics,'' Accessed on Jan 21, 2022.

\bibitem{SilvaAHWBSF2019}
W.~M. C.~d. Silva, J.~L. Andersen, M.~T. Holanda, M.~E. M.~T. Walter, M.~M.
  Brigido, P.~F. Stadler, and C.~Flamm, ``Exploring plant sesquiterpene
  diversity by generating chemical networks,'' {\em Processes}, vol.~7, no.~4,
  2019.

\bibitem{AndersenFMS2015}
J.~L. Andersen, C.~Flamm, D.~Merkle, and P.~F. Stadler, ``In silico support for
  eschenmoser’s glyoxylate scenario,'' {\em Israel Journal of Chemistry},
  vol.~55, no.~8, pp.~919--933, 2015.

\bibitem{AndersenMR2019}
J.~L. Andersen, D.~Merkle, and P.~S. Rasmussen, ``Combining graph
  transformations and semigroups for isotopic labeling design,'' {\em Journal
  of Computational Biology}, vol.~27, no.~2, pp.~269--287, 2020.

\bibitem{AndersenFFKMS2018}
J.~L. Andersen, R.~Fagerberg, C.~Flamm, R.~Kianian, D.~Merkle, and P.~F.
  Stadler, ``Towards mechanistic prediction of mass spectra using graph
  transformation,'' {\em MATCH, Communications in Mathematical and in Computer
  Chemistry}, vol.~80, no.~3, pp.~705--731, 2018.

\bibitem{hcn:13}
J.~L. Andersen, T.~Andersen, C.~Flamm, M.~M. Hanczyc, D.~Merkle, and P.~F.
  Stadler, ``Navigating the chemical space of hcn polymerization and
  hydrolysis: Guiding graph grammars by mass spectrometry data,'' {\em
  Entropy}, vol.~15, no.~10, pp.~4066--4083, 2013.

\bibitem{flow}
J.~L. Andersen, C.~Flamm, D.~Merkle, and P.~F. Stadler, ``Chemical
  transformation motifs --- modelling pathways as integer hyperflows,'' {\em
  IEEE/ACM Transactions on Computational Biology and Bioinformatics}, vol.~16,
  pp.~510--523, March 2019.

\bibitem{MetaCyc2019}
R.~Caspi, R.~Billington, I.~M. Keseler, A.~Kothari, M.~Krummenacker, P.~E.
  Midford, W.~K. Ong, S.~Paley, P.~Subhraveti, and P.~D. Karp, ``{The MetaCyc
  Database of Metabolic Pathways and Enzymes - a 2019 Update},'' {\em Nucleic
  Acids Research}, vol.~48, pp.~D445--D453, 10 2019.

\bibitem{AndersenFMS2013}
J.~L. Andersen, C.~Flamm, D.~Merkle, and P.~F. Stadler, ``{Inferring Chemical
  Reaction Patterns Using Rule Composition in Graph Grammars},'' {\em J. Syst.
  Chem.}, vol.~4, no.~1, p.~4, 2013.

\end{thebibliography}

{
	\onecolumn

\pagenumbering{Roman}
\appendix

\setcounter{figure}{0}
\renewcommand{\thefigure}{S\arabic{figure}}

\section{Supporting Information}

\subsection{Graph Transformation for Chemistry}
\label{app:gt_in_chem}

In this section we introduce the graph transformation concepts necessary for mathematically rigorous treatment of overlay graphs and substrate rules. While formal, the definitions utilize chemical examples. For a purely formal, category theoretical treatment, please refer to \hyperref[app:category_theory]{Section~\ref{app:category_theory}}.

The main text offers a simplified view on the notion of a graph-transformation rule. In practice, we utilize the double pushout formalism~\cite{HabelMP2001, AndersenFMS2016}, in which the correpondence between atoms in $L$ and $R$ is established by means of a graph $K$ and two maps from (the atoms in) $K$ to (atoms in) $L$ and $R$. Formally, a rule is a span $p = (L \xleftarrow{l}{} K\xrightarrow{r}{} R)$, where $L$ and $R$ are the left (in) and right (out) graphs, respectively. $K$ is the invariant graph containing elements common to $L$ and $R$ as specified by the injections $l$ and $r$. What is in $L$ but not in $K$ vanishes and what is in $R$ but not in $K$ appears.

\begin{figure}[!ht]
\begin{center}
\includegraphics[scale=.8]{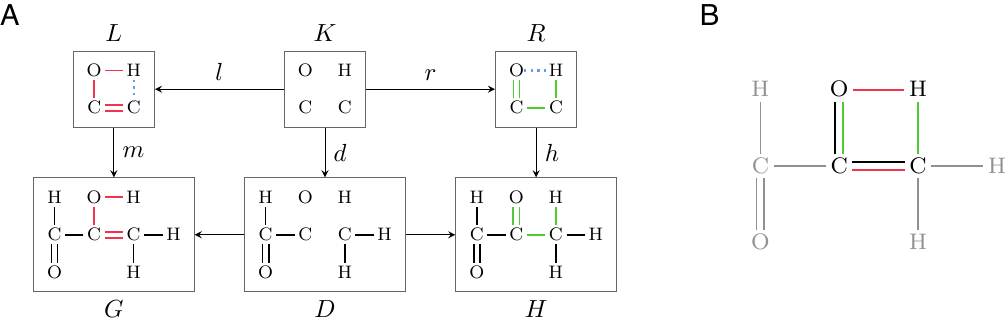}
\caption[Rule]{{\bf A:} The example rule describing a keto-enol tautomerization, applied to a 2-hydroxyacrylaldehyde in full formalism. To hint at the actual mappings $l$ and $r$, we use a spatial layout of $K$ in which the position of an atom is meant to suggest that it is mapped to the atom at the same (relative) position in $L$ and $R$. {\bf B:} The ITS of the reaction.}
\label{fig:rule_dpo}
\end{center}
\end{figure}

The application of a rule $p$ to a mixture $G$ occurs in two steps, illustrated in \hyperref[fig:rule_dpo]{Figure~\ref{fig:rule_dpo}A}. First, an embedding or match, $m$, of $L$ in $G$ must be found. Second, given $m$, those parts of $G$ that are in $L$ but not in $K$ (and thus meant to vanish) are removed to be replaced by those parts in $R$ that are not in $K$ (thus meant to appear), yielding the new mixture $H$. In the standard approach suitable for chemistry, punching a hole in $G$ is subject to some restrictions, such as avoiding the creation of dangling bonds. Absent dangling bonds, some information is needed for filling the hole in $G$ with $R$. This information comes from $K$ and the associated maps $l$ and $r$, which specify which left-over parts at the location of the match $m$ are meant to be the same parts in both $L$ and $R$. Given this identification, $R$ can be properly \enquote{glued} in place. All this is captured by the commuting diagram shown in \hyperref[fig:rule_dpo]{Figure~\ref{fig:rule_dpo}A}, where
the top row is the rule, $p$, and $G$ in the bottom row is the given mixture. Punching the hole in $G$ amounts to constructing $D$, which together with $K$ provides the instruction for gluing $R$ in place.

We write $\derivation{G}{H}{p, m}$ for the transformation of $G$ into $H$ by rule $p$ at the location of the match $m$ and refer to it as a \emph{direct derivation}. It is important to keep in mind that $G$ and $H$ are graphs representing fully specified molecules. Typically $G$ comprises the reactants of a particular reaction, but $G$ could also represent a large mixture within which a reaction takes place at the location indicated by $m$. (A direct derivation is really just a state transition, but we soon need more vocabulary.) If by virtue of the action of $p$ the new graph $H$ comes to satisfy the left pattern of rule $q$, we can view the application of $p$ as enabling, or causing, an application of $q$. In this way, step by step, a sequence of direct derivations $\derivation{G}{H}{p_1,m_1}\equiv \derivation{H_1}{H_2}{p_2, m_2},\dots , \derivation{H_{n-1}}{H_n}{p_n, m_n}$ represents a causal trajectory of transformations converting educts $G$ into products $H_n$ while at the same time mapping atoms in $G$ to atoms in $H_n$. We refer to the whole sequence as a \emph{derivation}.

On top of the non-bond constraints introduced in \hyperref[sec:method]{Section~\ref{sec:method}} of the main text, we need another syntactical construct specific to chemical applications. An atom that changes state, such as charge, should preserve its identity. In other words, a change of state should not be interpreted as the deletion of an atom in one state and the re-creation of an atom of the same type in the new state. We use a bullet in $K$ to indicate that the identity of the corresponding atom is preserved but that it acquires a new label to reflect the change of state, such as going from \ch{N} to \ch{N+}. An example of such untyped, or better yet unlabeled, vertices is available in~\hyperref[fig:formal_charges]{Figure~\ref{fig:formal_charges}A} making use of the spontaneous breakup of a \ch{phenyl-N-(sulfonatooxy)methanimidothioate} retrieved from the \href{https://metacyc.org/META/NEW-IMAGE?type=REACTION&object=RXN-12027}{MetaCyc} database~\cite{MetaCyc2019}.

\begin{figure}[!ht]
\begin{center}
\includegraphics[scale=.8]{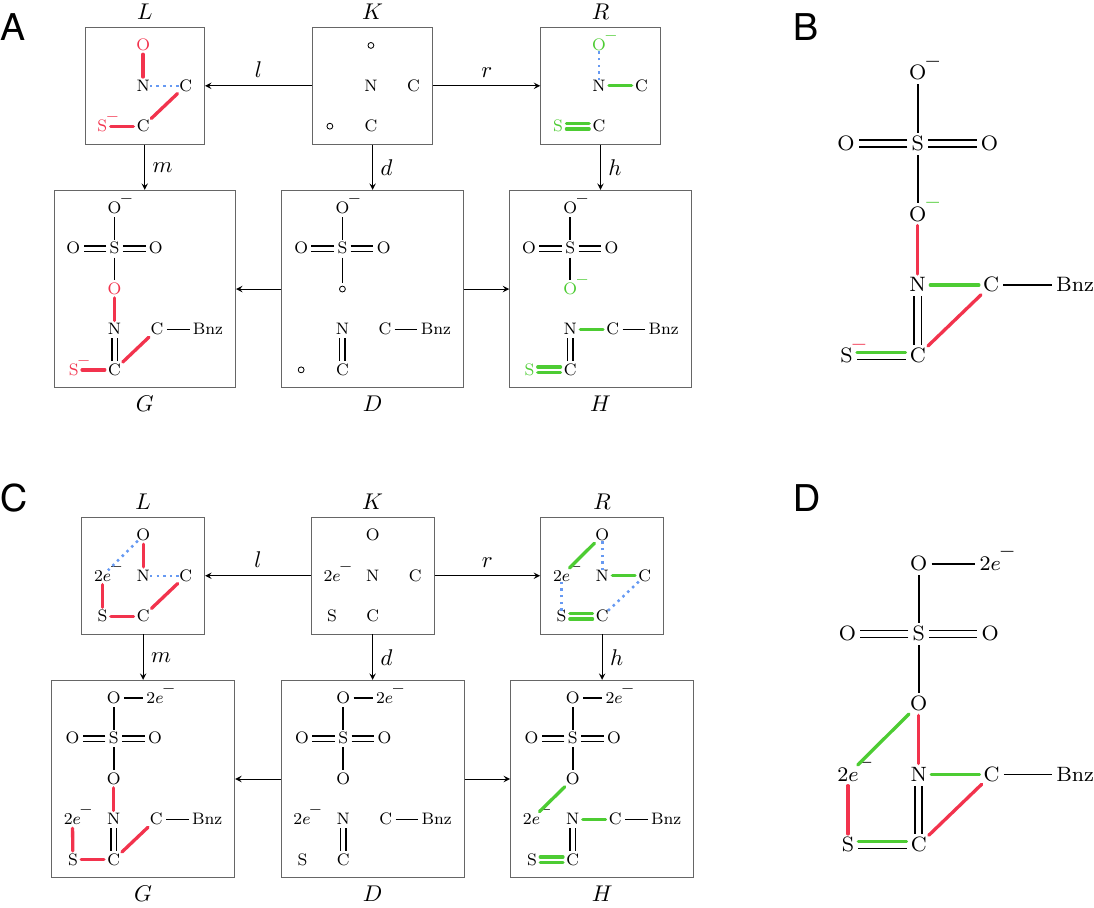}
\caption{
A rule applied to phenyl-N-(sulfonatooxy)methanimidothioate, representing its spontaneous breakup into a benzylisothiocyanate and a sulfate. This rule modifies formal charges, requiring the affected atoms to appear untyped in the invariant graph $K$. The ITS of the reaction is depicted on the right hand side.}
\label{fig:formal_charges}
\end{center}
\end{figure}

While both non-bond constraints and unlabeled vertices in $K$ are sound~\cite{AndersenFMS2016,AndersenFMS2018,BehrKAM2021}, at present, there's a lack of comprehensive support within graph transformation frameworks. As addressing this deficiency is not our primary concern, it should be noted that our implementation circumvents the issue by exploiting the structure of our input data, and does not generalize beyond our application domain.

Note that the ITS in \hyperref[fig:formal_charges]{Figure~\ref{fig:formal_charges}B} is not cyclical, even though the reaction is elementary. This is due to the changes in formal charges acting as sources and sinks for electron pairs. One can recover an ITS that is a loop by considering a virtual entity representing a free electron pair, as illustrated in \hyperref[fig:formal_charges]{Figure~\ref{fig:formal_charges}C and D}. This representation would not require label changes, but relies on artificial constructs, such as the explicit transfer of the free electron pair from the sulfur atom to the oxygen atom.

\subsubsection{Rule Composition}
\label{app:rule_comp}

Our focus is on mechanisms of enzymatically catalyzed reactions. The significance of a mechanism consists in providing a series of elementary steps leading from educts to products,  as well as an atom map, allowing a representation as a derivation -- sequence of rule applications, each corresponding to a step of the mechanism. We thus do not aim to describe the composition of rules in full generality~\cite{AndersenFMS2013, AndersenFMS2018}, but focus on the specific case of rule composition along a derivation.

We illustrate rule composition for the base case of two steps in which first rule $p_1$ is applied to the initial mixture, followed by an application of rule $p_2$. The case of $n$ steps is simply an iteration of the base case. Conceptually, the left pattern $L$ of the composite rule $p$ must collect the conditions (the graph parts) demanded by the left pattern of $p_1$ plus those conditions of the left pattern of $p_2$ that have not been created by the action of $p_1$. Likewise, the right pattern $R$ of $p$ must contain the right pattern of $p_2$ plus those parts of the right pattern of $p_1$ that have not been modified by $p_2$. The conversion of this specification into a formal construction makes use of standard operations on graphs, which we briefly explain to provide a self-contained presentation using as an example the first two steps of the mechanism in the main text, \hyperref[fig:mech_example]{Figure~\ref{fig:mech_example}}.

\begin{figure}[!ht]
\begin{center}
\includegraphics[scale=.8]{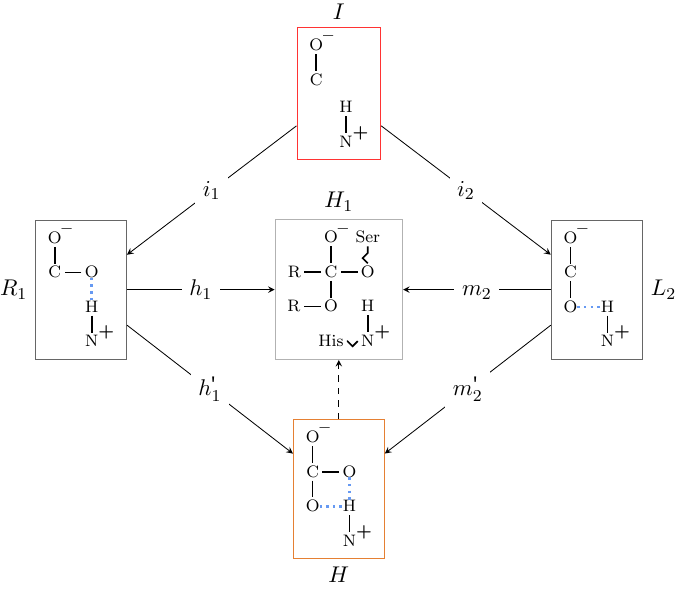}
\caption[Gluing]{Diagram showing the construction of the graph $H$ as detailed in the text. Note that $R_1$ and $L_2$ look the same in this case. Yet, the gluing instruction $I$ does not mention the neutral oxygen atoms, which means that their instances in $R_1$ and $L_2$ are not equivalent and thus not be identified in the merging of $R_1$ and $L_2$. $H$ can be viewed as a union of $R_1$ and $L_2$ based on the identifications made by $I$ with the maps $i_1$ and $i_2$. In this example $R_1$ and $L_2$ are the right and left patterns of the rules representing the first and second steps, respectively, of the triacylglycerol lipase mechanism. The source of information captured in $I$ and the maps $i_1$ and $i_2$ is the gray graph $H_1$ in the middle, which is the fully specified molecular graph (the mixture) just after the first step has been applied (with $h_1$ showing the location of $R_1$) and prior to the second step, showing the location $m_2$ where it will be applied. This information is part of \enquote{knowing the mechanism} and distinguishes the two \ch{C-O} components. The mechanism is necessary to define the gluing instruction $I$.}
\label{fig:h_pushout_hyperdiagram}
\end{center}
\end{figure}

The key operation is the gluing together of two graphs, which occurs by declaring certain nodes and links in one graph as equivalent to certain nodes and links in the other, then identifying (merging) the equivalent graph elements to combine the two graphs. The declaration of equivalence amounts to a \enquote{gluing instruction}. It must respect type; so we cannot identify an \ch{O} with a \ch{C} or a single bond with a double bond.

\begin{figure}[!ht]
\begin{center}
\includegraphics[scale=.8]{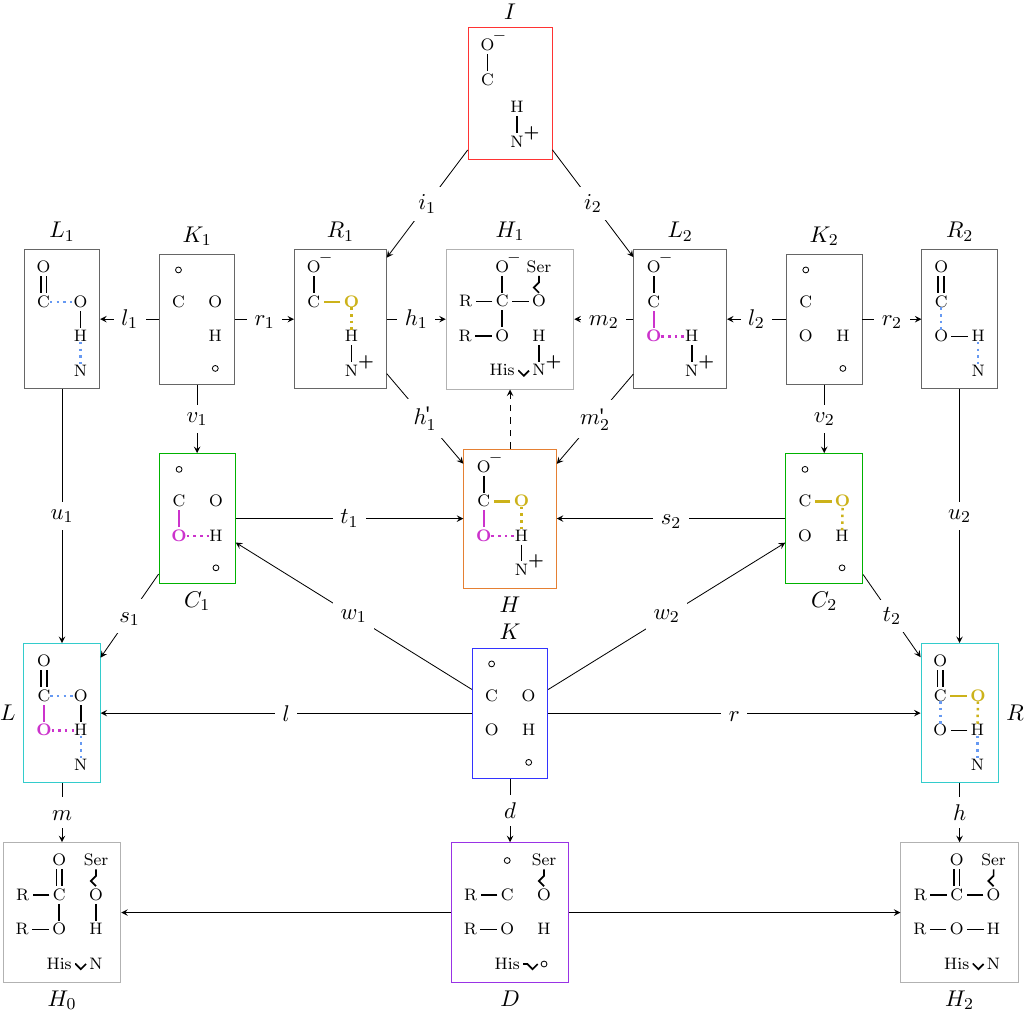}
\caption[Composition]{Diagram with the sequence of constructions needed to compose the two rules $p_1 = (L_1 \xleftarrow{l_1}{} K_1 \xrightarrow{r_1}{} R_1)$ and $p_2 = (L_2 \xleftarrow{l_2}{} K_2 \xrightarrow{r_2}{} R_2)$. The information in $I$ comes from the mechanism, i.e.~the result ($h_1$) of the application of $p_1$ in a concrete mixture and the embedding ($m_2$) of the left pattern of $p_2$ in preparation for the subsequent application of $p_2$ (gray structures). This information allows the construction of $H$, which contains the graphical elements that must be merged with $L_1$ and $R_2$ to obtain the left and right patterns $L$ and $R$, respectively, of the composite rule. After $H$, $C_1$ ($C_2$) is constructed, which in turn allows the construction of $L$ ($R$) as detailed in the text. The commutative property underlying these manipulations enables their automatic execution.}
\label{fig:composition_hyperdiagram}
\end{center}
\end{figure}
\afterpage\clearpage

For example, the graph $I$ at the top of \hyperref[fig:h_pushout_hyperdiagram]{Figure~\ref{fig:h_pushout_hyperdiagram}} is a gluing instruction asserting that the \ch{O^{-}-C} part is the same in both $R_1$ and $L_2$; likewise for \ch{H-N^{+}}. This correspondence is made explicit by the mappings $i_1$ and $i_2$, which are given here implicitly by the layout of the atoms. Note that in this example $R_1$ and $L_2$ are isomorphic, which illustrates why a gluing instruction is necessary: the fact that $I$ does \emph{not} mention a neutral \ch{O} means that the (uncharged) oxygen atoms in $R_1$ and $L_2$ are not to be identified; they are distinct instances of \ch{O}. Given $I$ and the maps $i_1$ and $i_2$, the graphs $R_1$ and $L_2$ can be glued together uniquely yielding $H$. If $R_1$ and $L_2$ contained multiple instances of, say, \ch{O^{-}-C}, the mappings $i_1$ and $i_2$ would determine which one corresponds to which. This raises the question about the choice of $i_1$ and $i_2$ when there are multiple possibilities. The answer comes from knowing the actual mechanism, given by the execution of the rules in the concrete setting of a mixture of reactants. The notion of mechanism includes the graph $H_1$, which is the intermediate mixture, along with the embedding maps $h_1$ and $m_2$ telling us where rule $p_1$ acted and where rule $p_2$ is supposed to act. In the present case, the mechanism distinguishes the two (uncharged) oxygen atoms. The \ch{O} mentioned in $R_1$ comes from the serine, which binds the substrate to the enzyme as a result of step 1; whereas the \ch{O} in the isomorphic $L_2$ belongs to the substrate, from which it will be severed in step 2. Hence, the composition of two rules, $p_1 = (L_1 \xleftarrow{l_1}{} K_1 \xrightarrow{r_1}{} R_1)$ and $p_2 = (L_2 \xleftarrow{l_2}{} K_2 \xrightarrow{r_2}{} R_2)$ depends on where in the mixture $p_1$ and $p_2$ act. This location is part of the mechanism and is given by $H_1$, which contains the result of $p_1$ (i.e.~$R_1$) and satisfies the requirements of $p_2$ (i.e.~$L_2$). We thus annotate the composition operator with the location information: $p = p_1 \mathbin{\rulecomp_{H_1}} p_2$.

We next use the gluing construction to determine the left ($L$) and right ($R$) pattern of the composite rule, as illustrated in \hyperref[fig:composition_hyperdiagram]{Figure~\ref{fig:composition_hyperdiagram}}. The orange-framed graph is $H$ of \hyperref[fig:h_pushout_hyperdiagram]{Figure~\ref{fig:h_pushout_hyperdiagram}}, which is the gluing of $R_1$---the structure created by $p_1$---with $L_2$---the structure required by $p_2$. Thus, $H$ contains both the graph fragments requested by $L_2$ but not delivered by $p_1$ and the modifications caused by $p_1$ that survive the action of $p_2$ (because their presence is ignored by $L_2$). We only need to tease these contributions apart and glue them to $L_1$ and $R_2$ to obtain $L$ and $R$ of the composite rule. 

Note that $H$ can also be viewed as the gluing of $R_1$ with an unknown graph $X$ using the gluing instructions provided by $K_1$. The difference between $R_1$ and $H$, highlighted in magenta, consists of the fragment [\ch{\bond{single}O\bond{dotted}}] comprising the oxygen atom of the substrate, its single bond to the carbon of the substrate and the non-bond constraint with the \ch{H} of the protonated histidine. This fragment must therefore occur in $X$. The unknown graph we seek is given by $C_1$, which is $H$ from which $R_1$ has been removed while keeping the fragments indicated by $K_1$. With $C_1$ in hand, we can determine $L$ as the gluing of $L_1$ with $C_1$. We now see that the [\ch{\bond{single}O\bond{dotted}}] fragment, which is part of the structural condition required by $p_2$ and not delivered by $p_1$, has been included as a requirement in the left pattern $L$ of the composite rule. Using the same reasoning we transport the other [\ch{\bond{single}O\bond{dotted}}] fragment, highlighted in yellow, to the $R$-side. This fragment comprises the oxygen of the serine, its single bond to the carbon of the substrate and the non-bond constraint with the protonated histidine. The same logic is deployed to obtain the invariant graph $K$ of the composite rule as the gluing instruction that directs $C_1$ and $C_2$ to be combined into a graph compatible with $H$.

The composite rule $p=(L \xleftarrow{l}{} K \xrightarrow{r}{} R)$ can be applied to a mixture $H_0$ in which it induces the direct derivation (the state change) $\derivation{H_0}{H_2}{p, m}$, as shown at the bottom of \hyperref[fig:composition_hyperdiagram]{Figure~\ref{fig:composition_hyperdiagram}}, without a need for the intermediate stage $H_1$. This procedure can now be iterated for all steps of a mechanism to construct the composite rule of the overall reaction. For the mechanism from main text \hyperref[fig:mech_example]{Figure~\ref{fig:mech_example}}, this composite rule $p$ can be found as $p = (L \xleftarrow{l}{} K\xrightarrow{r}{} R)$ in the second row of \hyperref[fig:substrate_hyperdiagram]{Figure~\ref{fig:substrate_hyperdiagram}}.

We can now see that the partial catalog of changes the rule composition retains in place of the full causal information in the mechanism is largely recorded withing the invariant graph $K$ of the composite rule. (i) The doubly-bound \ch{O} and the \ch{N} must have undergone a transient change, because $K$ has no type specification at those nodes and yet they are present in both $L$ and $R$. (ii) Likewise, the double bond is in both $L$ and $R$, but not in $K$; hence, it must have been transiently destroyed and then reconstituted. (iii) The \ch{N} and \ch{H} must have been temporarily bound to each other, because the non-bond constraint is present in $L$ and $R$, but not in $K$. (iv) One \ch{C-O} bond has been cleaved and a different one has been created. (v) Similarly, one \ch{O-H} bond was lost and another was gained. The latter two changes are made explicit as differences between $L$ and $R$. The point is that the composite rule also preserves transitory changes, because, by construction, $K$ no longer specifies just the parts that are the same in $L$ and $R$, but rather the parts that stay the same \emph{throughout} the mechanism.

The composite rule represents the reaction mechanism in a compact form. The combined graph, or \ourITSname{} then contains all atoms, bonds and explicit non-bond constraints that appear on either side of the composite rule. Since the rule also records transient changes, we use a color scheme to annotate the change of type that various parts experience as they pass from $L$ to $R$. A bond (or atom) is shown in green when the net change across the reaction is a formation and in red when the net change is an elimination. Bonds and atoms whose original state is eliminated and then re-introduced are shown in purple. For non-bond constraints, which flag an absence, we color transient changes in blue to emphasize that an introduction is followed by an elimination (the opposite of purple transients).

The example used to illustrate the process of rule composition included steps that covered only the reaction center, that is, those parts of a molecular graph that are subject to modifications. This is why the invariant graph $K$ does not show any parts that remain constant throughout the mechanism, except for atoms. In practice, a rule may include molecular graph fragments that are known to constitute a condition for reaction but are not altered by the mechanism, such as heteroatoms or functional groups that modify the electronic density or substituents that act as steric constraints. These truly constant parts (as opposed to the transiently-modified \enquote{constant} parts) show up in $L$, $R$ and in $K$. They are inherited by $K$ because they would occur in $K_1$ and $K_2$, and thus in $C_1$ and $C_2$ from which $K$ is constructed.

\subsubsection{Substrate Rules}
\label{app:substrate_rule}

Finally, we present the formal version of the substrate rule inference. To construct the substrate rule associated with a catalytic process, we proceed as illustrated in \hyperref[fig:substrate_hyperdiagram]{Figure~\ref{fig:substrate_hyperdiagram}}. In essence, we need to remove the parts representing the catalyst from the composite rule $L \xleftarrow{l}{} K \xrightarrow{r}{} R$, shown in the second row of \hyperref[fig:substrate_hyperdiagram]{Figure~\ref{fig:substrate_hyperdiagram}}. (Note that the composite rule is not the one of \hyperref[fig:composition_hyperdiagram]{Figure~\ref{fig:composition_hyperdiagram}}, which resulted from composing only step 1 and step 2, but the rule resulting from the composition of all four steps of the full mechanism of the main text \hyperref[fig:mech_example]{Figure~\ref{fig:mech_example}}.) The graph $C$ is our choice of helper molecules. The graph $C$, alongside the mappings $c_1$ and $c_2$, identifies the catalyst in the educt mixture $E$ and the product mixture $P$. By using the empty graph as a gluing instruction we effectively remove the graph parts of $C$ from $E$ and $P$ to yield $E'$ and $P'$ (top row), respectively. $L'$ is the intersection of $L$ and $E'$; similarly for $R'$. To obtain the graph $K'$ for the substrate rule we first construct $N_1$, which is the intersection of $K$ (whose map $l$ tells us what stays invariant in $L$) and $L'$, since $L'$ is in $L$. Similarly for the $R'$ side to get $N_2$. $K'$ then is the intersection of $N_1$ with $N_2$.

\begin{figure}[!ht]
\begin{center}
\includegraphics[scale=.8]{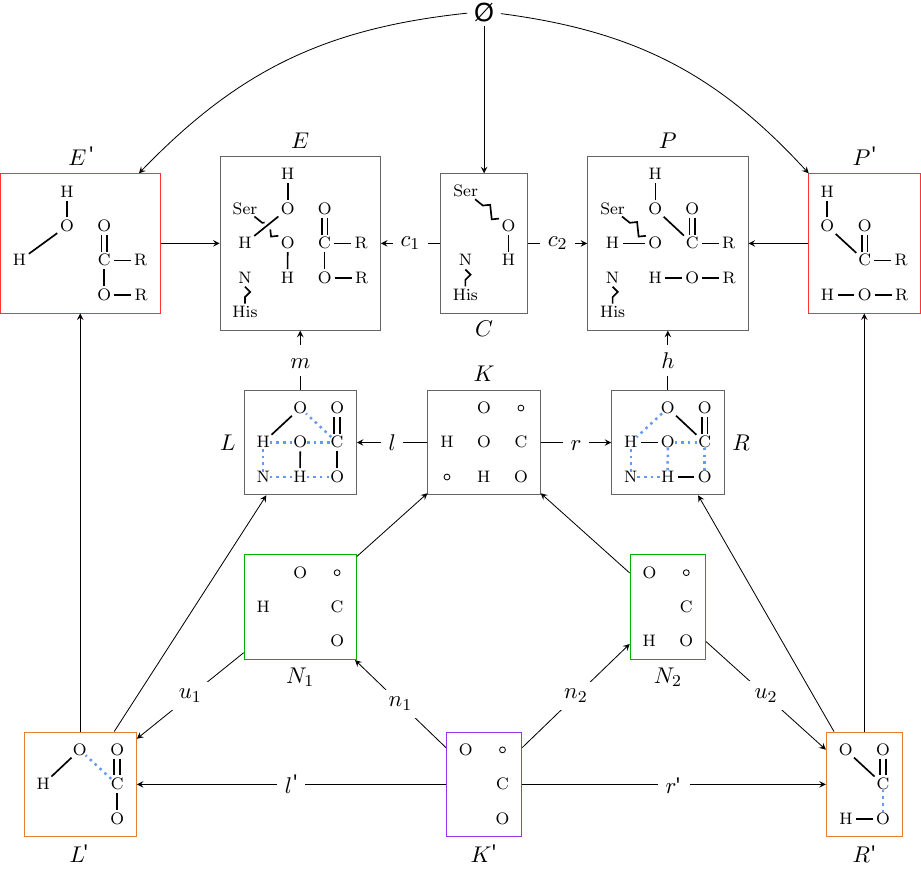}
\caption[Substrate rule]{Diagram for constructing  the substrate rule. The colors indicate the progression of graph constructions: red first, then orange, then green and finally purple. See text for details.}
\label{fig:substrate_hyperdiagram}
\end{center}
\end{figure}

The choice of $C$, $c_1$ and $c_2$, is subject to to the obvious constraint that $C$ must be a subgraph of both $E$ and $P$. The construction of the substrate rule and the applicability of the rule, however, pose further constraints. In particular, the constraints arise due to the possibility of introducing dangling bonds. The potential to introduce dangling bonds is twofold. During the construction of $E'$ and $P'$, and during the application of the substrate rule to the original mixture.

The case of $E'$ and $P'$ is relatively simple. Any vertex in $C$ is being removed from $E$ to obtain $E'$. All edges such a vertex has in $E$ have to be reflected in $C$, including their endpoints. The situation is the same for $P'$ and $P$, constraining $C$ to correspond to connected components, or fully specified molecules, in $E$ and $P$. This constraint is in general not an issue for the substrate rules, as they are constructed by explicitly removing fully specified molecules. However, as we demonstrated with the notion of catalytic complement OG, one may be interested in considering only specific functional groups (\enquote{catalytic sites}) for inclusion in the graph $C$. 

For example, recall the lower left quadrant in \hyperref[fig:311_comp]{Figure~\ref{fig:311_comp}A} of the main text, which depicts an overlay graph of M-CSA \mcsaentry{748}. In this case, the electron source is a nitrogen of an amine group that belongs to the substrate. The intra-molecular connection between the carboxylic ester carbon and the catalytic amine group consists of a single carbon atom, \hyperref[fig:catalytic_complement]{Figure~\ref{fig:catalytic_complement}}, but is abstracted away (as symbolized by the wavy line in the main text \hyperref[fig:311_comp]{Figure~\ref{fig:311_comp}A}) because our minimalistic rules and OGs only include bonds and atoms that directly participate in the reaction.The catalytic nitrogen of the substrate thus appears as a disconnected graphical component. In case the connecting carbon is included in the rule and the OG, one has to ensure that the \ch{C-N} bond to the amine nitrogen is removed during the construction of the catalytic complement OG, alongside the nitrogen itself. The \ch{C-N} bond would then necessarily appear in the graph $C$. Preventing the erasure of the carbon atom from $E'$ and $P'$ is achieved by adding it to the empty graph at the top of \hyperref[fig:substrate_hyperdiagram]{Figure~\ref{fig:substrate_hyperdiagram}}.

\begin{figure}[!ht]
\begin{center}
\includegraphics[scale=1]{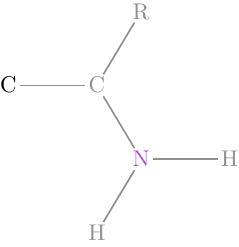}
\caption{%
The amine group of the tRNA bound D-amino acid substrate, acting as a catalyst of (M-CSA \mcsaentry{748}), in relation to the carboxylic ester carbon (black).}
\label{fig:catalytic_complement}
\end{center}
\end{figure}

The second instance occurs after the substrate rule has been obtained and restricts its applicability to the original mixture. The restriction becomes relevant if graph $L'$ has more vertices than $K'$, meaning that the substrate rule deletes atoms, which have to show up outside the substrate in the catalytic part of the reaction. These \enquote{catalytic atoms} are identified by choosing the maps $c_1$ and $c_2$, which define an atom map from the catalytic parts of $E$ to the catalytic parts of $P$. However, the atom map defined by $c_1$ and $c_2$ could be at odds with the atom map defined by the mechanism and represented by the morphisms $l$ and $r$ of the associated composite rule. This could thwart the applicability of the substrate rule. To avoid this situation, the following must be ensured: all bonds of an atom in $E$ that is mapped by $c_1$ and $c_2$ to an atom in $P$ different than the atom to which it is mapped by $l$ and $r$ must be represented in $L$. In other words, if an atom that is part of a catalytic moiety becomes part of another catalytic moiety (hence is being created or deleted by the substrate rule), then all of its bonds have to be accounted for in the composite rule. This can usually be achieved by choosing $c_1$ and $c_2$ in a way that maximizes their overlap with the atom map of the mechanism. There are interesting exceptions, however, which we explore in greater detail in \hyperref[app:catalyst_ambiguity]{Section~\ref{app:catalyst_ambiguity}}.

\subsection{Graph Transformation in Category Theory}
\label{app:category_theory}

This material reiterates the constructions of the composite rule and substrate rule from the previous section in full categorical formalism. At least an intuitive understanding of the categorical notions of span, pullback, pushout and pushout complement is needed.

First, we revisit the composition of two rules $p_i = L_i \xleftarrow{l_i} K_i \xrightarrow{r_i} R_i$ equipped with direct derivations $\derivation{H_{i-1}}{H_i}{p_i, m_i}$ for $i\in \{1,2\}$. \hyperref[fig:h_pushout_diagram]{Figure~\ref{fig:h_pushout_diagram}} shows the construction of the shared subgraph $H$ via a pullback from the co-span $R_1 \xrightarrow{h_1} H_1 \xleftarrow{m_2} L_2$ and a subsequent pushout. Note that $h_1$ and $m_2$ come from the direct derivations as illustrated in \hyperref[fig:rule_dpo]{Figure~\ref{fig:rule_dpo}A}.

\begin{figure}[!ht]
\begin{center}
\includegraphics[scale=1]{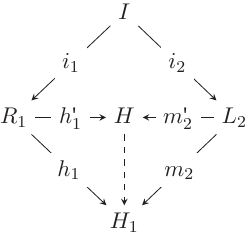}
\caption{
Commutative diagram showing the construction of the graph $H$, as a pullback from the co-span $R_1 \xrightarrow{h_1} H_1 \xleftarrow{m_2} L_2$ and a subsequent pushout of the obtained span $R_1 \xleftarrow{i_1} I \xrightarrow{i_2} L_2$. The dashed arrow indicates the unique commuting morphism embedding $H$ into $H_1$.}	
\label{fig:h_pushout_diagram}
\end{center}
\end{figure}

The graph $H$ (which is part of \enquote{knowing the mechanism}) is an input to the rule composition. The composition itself is carried out in three steps, \hyperref[fig:composition_diagram]{Figure~\ref{fig:composition_diagram}}:
\begin{enumerate}
    \item $C_1$ and $C_2$ are constructed as pushout complements of $K_1 \xrightarrow{r_1} R_1 \xrightarrow{h_1'} H$ and $K_2 \xrightarrow{l_2} L_2 \xrightarrow{m_2'} H$, respectively.
    \item $L$ and $R$ are obtained as pushouts of the spans $L_1 \xleftarrow{l_1} K_1 \xrightarrow{v_1} C_1$ and $R_2 \xleftarrow{r_2} K_2 \xrightarrow{v_2} C_2$.
    \item The graph $K$ is produced as a pullback of the co-span $C_1 \xrightarrow{t_1} H \xleftarrow{s_2} C_2$.
\end{enumerate}

The rule $p=L\xleftarrow{l}K \xrightarrow{r} R$ is completed by taking $l = s_1 \funccomp w_1$ and $r = t_2 \funccomp w_2$. Finally, the composite derivation $\derivation{H_0}{H_2}{p, m}$ is such that $m\funccomp u_1$ commutes with $m_1$ and $h\funccomp u_2$ commutes with $h_2$.

\begin{figure}[!ht]
\begin{center}
\includegraphics[scale=1]{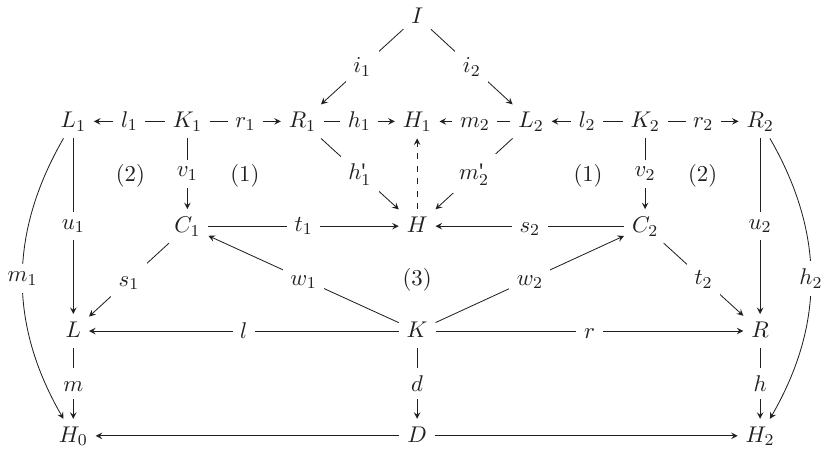}
\caption{Commutative diagram for rule composition of two rules
$p_1 = (L_1 \xleftarrow{l_1}{} K_1 \xrightarrow{r_1}{} R_1)$ and $p_2 = (L_2 \xleftarrow{l_2}{} K_2 \xrightarrow{r_2}{} R_2)$ with direct derivations $\derivation{H_0}{H_1}{p_1, m_1}$ and $\derivation{H_1}{H_2}{p_2, m_2}$ respectively. The resulting rule is $p= p_1 \mathbin{\rulecomp_{H_1}} p_2 = (L \xleftarrow{l}{} K \xrightarrow{r}{} R)$ with $l= s_1\funccomp w_1$ and $r = t_2 \funccomp w_2$. The numbers in parentheses denote the order in which the graphs are constructed by completing the commuting (sub)diagram.}
\label{fig:composition_diagram}
\end{center}
\end{figure}

The rule composition yielding also a direct derivation allows us to chain the compositions until the entire mechanism is represented in a single, composite rule.

To obtain the substrate rule, we once again rely on the categorical constructions of pullback and pushout complement. Given the monomorphisms $C \xrightarrow{c_1} E$ and $C \xrightarrow{c_2} P$, the construction of the substrate rule (\hyperref[fig:substrate_diagram]{Figure~\ref{fig:substrate_diagram}}) is the following multi-step process.

\begin{enumerate}
    \item The complement of $C$ in $E$ and the complement of $C$ in $P$, $E'$ and $P'$ respectively, are formally obtained as pushout complements of $\varnothing \xrightarrow{} C \xrightarrow{c_1} E$, respectively $\varnothing \xrightarrow{} C \xrightarrow{c_2} P$, where $\varnothing$ is the empty graph.
    \item The graphs $L'$ and $R'$ are constructed as pullbacks of the co-spans $E' \xrightarrow{} E \xleftarrow{m} L$ and $P' \xrightarrow{} P \xleftarrow{h} R$, effectively projecting $L$ and $R$ on the substrate subgraphs of $E$ and $P$, respectively.
    \item In the third step, two helper structures, $N_1$ and $N_2$, are obtained as pullbacks of the co-spans $L' \xrightarrow{} L \xleftarrow{l} K$ and $R' \xrightarrow{} R \xleftarrow{r} K$, respectively.
    \item Finally, $K'$ is constructed as a pullback of the co-span $N_1 \xrightarrow{} K \xleftarrow{} N_2$.
\end{enumerate}

\begin{figure}[!ht]
\begin{center}
\includegraphics[scale=1]{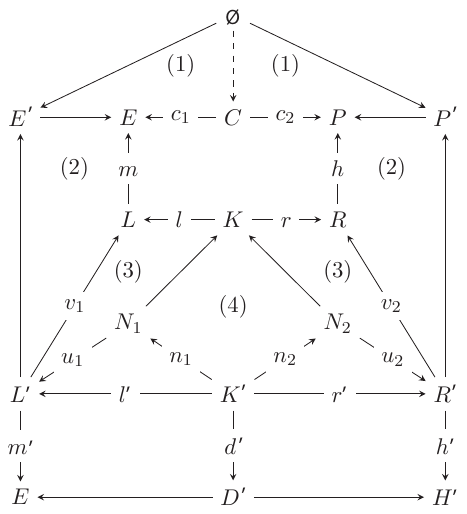}
\caption{%
Commutative diagram inferring the substrate rule using categorical constructions. $E'$ and $P'$ are the substrate part of the educt graph and product graph, respectively, constructed as pushout complements. The left and right graphs $L'$ and $R'$ of the substrate rule correspond to the shared part of $E'$ and $L$ and of $P'$ and $R$, respectively, constructed as pullbacks. Similarly, $K'$ corresponds to the shared part of $L'$, $R'$ and $K$. The numbers in parentheses denote the order in which the graphs are constructed.}
\label{fig:substrate_diagram}
\end{center}
\end{figure}

The substrate rule $L' \xleftarrow{l'} K' \xrightarrow{r'} R'$ is finalized by taking $l' = u_1 \funccomp n_1$ and $r' = u_2 \funccomp n_2$. The diagram in \hyperref[fig:substrate_diagram]{Figure~\ref{fig:substrate_diagram}} depicts the construction of the substrate rule for a given $C$, $c_1$ and $c_2$.

\subsection{Atom Exchange between Substrate and Catalyst}
\label{app:atom_exchange}

\begin{figure}[!ht]
\begin{center}
\includegraphics[scale=.8]{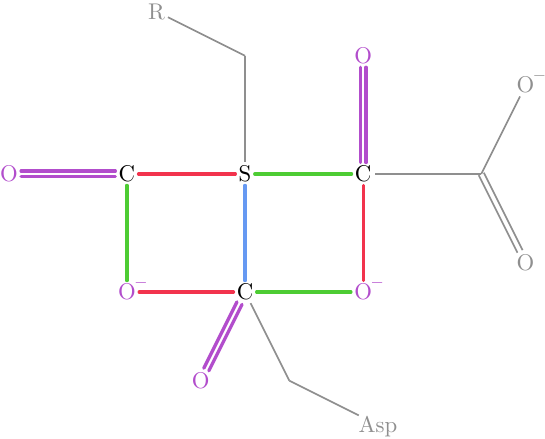}
\caption{%
An overlay graph of a formyl-CoA transferase reaction (M-CSA \mcsaentry{155}). The aspartate of the catalyst gives away its \ch{O-} to the formate product, indicated by the red bond at the bottom left. The aspartate is subsequently restored by an oxygen coming from the oxalate educt, green bond at the bottom right.}
\label{fig:oxygen_exchange}
\end{center}
\end{figure}

In the main text we mention the possibility that substrates and catalytic sites trade atoms. While this is very common for hydrogen atoms, as in the running example of the main text (M-CSA \mcsaentry{218}), heavier atoms, like oxygen, can also be exchanged, as shown in \hyperref[fig:oxygen_exchange]{Figure~\ref{fig:oxygen_exchange}}.

\subsection{Limits of the Rule Composition Approach}
\label{app:limits}

Rule composition amounts to coarse-graining a mechanism, because it cannot retain all the information contained in the ordered sequence of steps that constitutes a mechanism. Here we summarize what kind of information is lost by using as illustrative case the D-alanine transaminase reaction (M-CSA \mcsaentry{66}) whose automatically generated overlay graph is depicted in \hyperref[fig:transaminase]{Figure~\ref{fig:transaminase}A}. (The layout, however, was done by hand.)

\begin{figure}[!ht]
\begin{center}
\includegraphics[scale=.8]{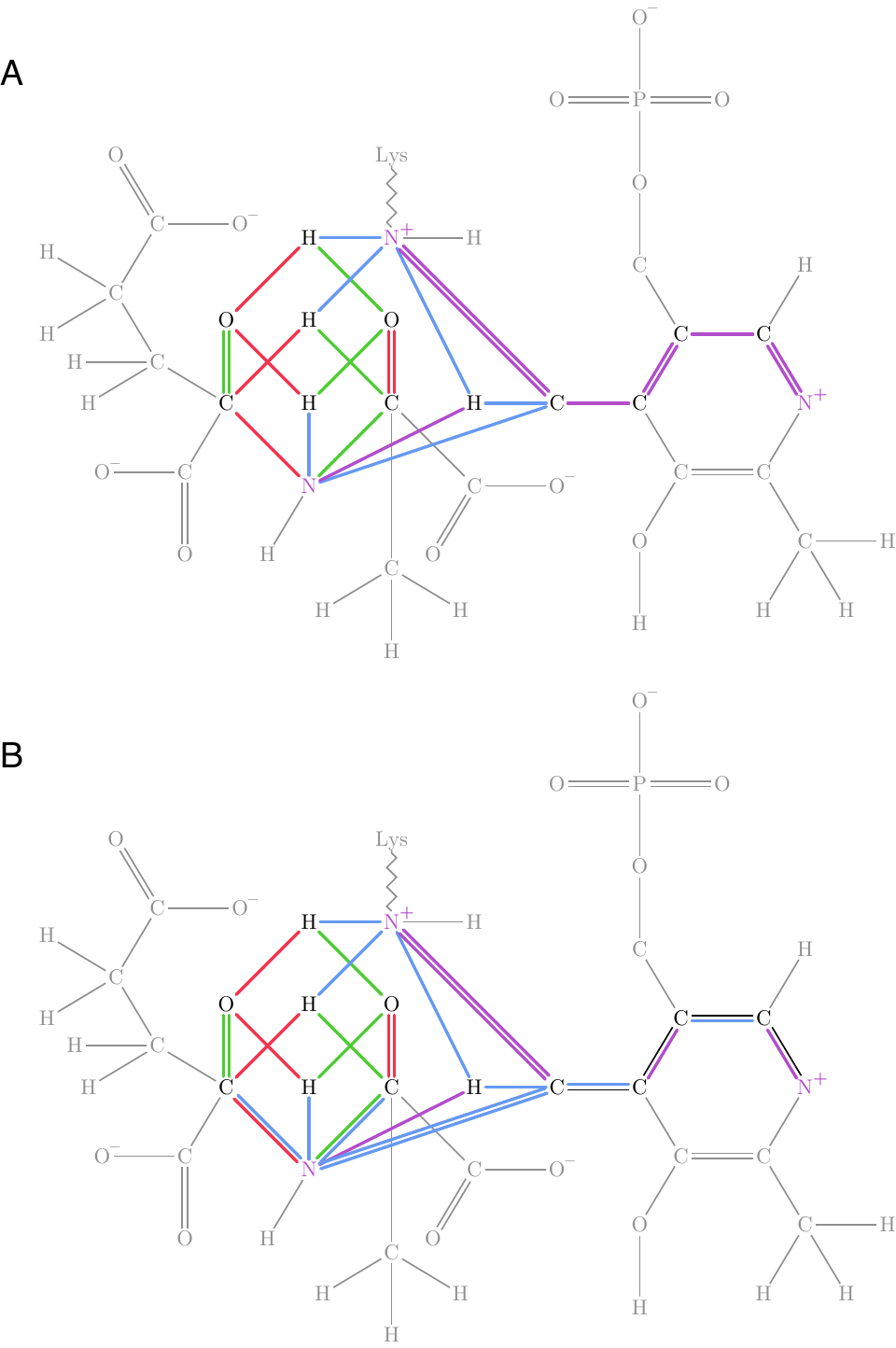}
\caption{%
{\bf A:} Overlay graph of the D-alanine transaminase reaction (M-CSA \mcsaentry{66}) resulting from rule composition. {\bf B:} An overlay graph of the same reaction created by hand to illustrate the full extent of information contained in the detailed mechanism.}
\label{fig:transaminase}
\end{center}
\end{figure}

The reaction mechanism extends over 12 steps and uses pyridoxal phosphate (PLP) as a cofactor. PLP starts out bound as a Schiff base to the amino end of a lysyl residue of the enzyme. It functions as both an electron sink through the positively charged nitrogen atom in its ring and as a transient carrier of an amino group, which it accepts from the glutamate substrate and transfers to the pyruvate with the assistance of an \ch{H2O} molecule. In this process, among other events, a bond between an \ch{H} and the \ch{N} of the substrate amino group is broken and later reconstituted (the purple \ch{N-H} edge in the overlay graph). In another step an electron pair is transiently parked at the ring nitrogen of PLP, which entails transiently flipping a number of double and single bonds (the purple chain \ch{C-C=C-C=N^+} in the overlay graph in \hyperref[fig:transaminase]{Figure~\ref{fig:transaminase}A}). 

We observe that the overlay graph of the composite rule correctly records the \ch{N-H} edge as transiently broken, but it also records the \ch{C-C} bonds in the purple chain as transiently broken, when, in fact, they transiently change from single to double bonds; similarly for the \ch{C=C} bonds in that same chain. 
Moreover, while it is the case that the \ch{C=N^+} bond between PLP and lysine is transiently broken, as reported in the overlay graph, it is also the case that in the process the same \ch{C} and \ch{N} are transiently linked through a single bond. The overlay graph only reports the \enquote{outer} or top-level modification and is blind to transient modifications nested within. In the same vein, the overlay graph does not record multiple transient modifications of the same kind between two atoms.

A non-bond constraint obviously carries no bond type information. The overlay graph can, therefore, only record a transient bond formation between PLP and the \ch{N} of the substrate (the blue \ch{C-N} edge), but cannot report that at some point it also is a double and not only a single bond. 

Finally, the overlay graph does not record transient changes that precede permanent changes, whether creation or dissolution, of a bond. For example, the \ch{C-N} single bond of the glutamate first becomes transiently a double bond before being broken; similarly, the newly created \ch{C-N} bond of the alanine product, resulting from the amino transfer to pyruvate, is first created as a double bond before being reduced to a single bond.

All this information is lost because nodes and edges of the invariant graph $K$ of a rule can only be typed by a label that matches these nodes and edges in both $L$ and $R$. Thus, if an edge is transiently modified from single bond to double bond, it shows up as a single bond in the $L$ and $R$ pattern of the composite rule, but as absent in the invariant graph $K$. From this, one can only deduce the existence of a transient modification, but not its kind. In the same vein, one cannot use $K$ to infer the multiplicity of the transient modifications, shall the same one occur multiple times, nor quantify or even detect that such an information loss occurred during the composition.

For comparison, \hyperref[fig:transaminase]{Figure~\ref{fig:transaminase}B} shows the overlay graph annotated by hand to aid in locating the incomplete information reported in panel A. It is possible to add type distinctions of the kind we used for state changes of atoms, but at that point one might as well forgo coarse-graining by dropping the composition construction altogether and instead work with the detailed sequence of mechanistic steps. This, on the other hand, might provide too much detail, limiting insight and hampering the automation of search. Especially with regard to the latter, the information lost by rule composition in no way limits retrieving reactions that are compatible with the mechanism implicit in the composite rule, which still accurately records the graphical requirements for a reaction, given its mechanism.

\subsection{Substrate Rules and the Notion of Catalyst}
\label{app:catalyst_ambiguity}

The substrate rule is derived from the composite rule by removing the parts that are responsible for catalysis in the reaction. The idea behind the substrate rule is to retain the condition of applicability of the mechanism while being as agnostic as possible about the chemical realization of the catalyst. In ultimate analysis, however, the idea of the substrate rule must be viewed as a heuristic, as it can infringe on the limits of the graph transformation framework. We briefly clarify this issue with two examples.

The first example is an isochorismate synthase reaction (M-CSA \mcsaentry{325}) in which the respective \ch{OH} of a water molecule becomes a new alcohol group of the substrate and an original alcohol group of the substrate is released as a water molecule (\hyperref[fig:isochorismate]{Figure~\ref{fig:isochorismate}}). According to the mechanism, the incoming and the outgoing water molecule share one hydrogen atom. This hydrogen atom is abstracted from the incoming water molecule, parked at a lysine residue and retrieved later in forming the outgoing water molecule during a clean-up step that restores the enzymatic state. The other hydrogen atoms of both the incoming and outgoing \ch{H2O} are the ones that travel with the \ch{OH} components. These hydrogen atoms never actively participate in the reaction (gray) and are therefore not represented in the left pattern of the composite rule or the associated substrate rule. When applying the substrate rule to a mixture, we need to be able to delete the oxygen atom indicated by an arrow in \hyperref[fig:isochorismate]{Figure~\ref{fig:isochorismate}A} from the substrate graph. Yet, there is no way of specifying this deletion within the double pushout framework, unless the bond to the passive hydrogen is included in the composite (and substrate) rule. Thus, the substrate rule cannot be applied despite there being only a single water molecule on the left and the right side of the reaction, qualifying it as a catalyst and making the choice of $c_1$ and $c_2$ in \hyperref[fig:substrate_diagram]{Diagram~\ref{fig:substrate_diagram}} unique. The key is that the \ch{H2O} on the left and the right consist of different atom identities, which makes the catalytic atom map defined by $c_1$ and $c_2$ (in which the \ch{H2O} on the left and the right appear identical) jar with the atom map of the mechanism.

\begin{figure}[!ht]
\begin{center}
\includegraphics[scale=.8]{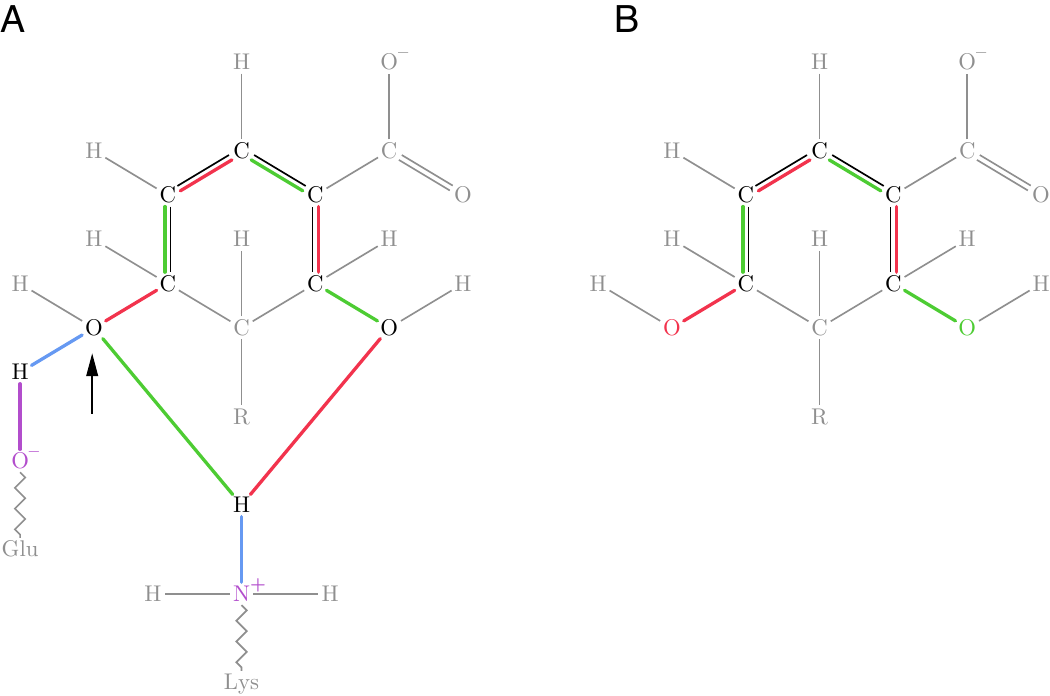}
\caption{%
{\bf A:} An overlay graph of an isochorismate synthase reaction (M-CSA \mcsaentry{325}). {\bf B:} The substrate overlay graph of the same reaction with \ch{H2O} in $C$. The rule is not applicable to the original educt mixture as the hydrogen bond (gray) of the red oxygen atom prevents its deletion.}
\label{fig:isochorismate}
\end{center}
\end{figure}

As a result, and for our purposes, the water molecule(s) cannot be declared catalytic in the sense of belonging to graph $C$ in \hyperref[fig:substrate_diagram]{Diagram~\ref{fig:substrate_diagram}}, unless the inactive hydrogen atoms are explicitly added to the invariant graph of the composite rule, thereby enabling their deletion and creation alongside the respective oxygen atoms.

The change of identity of catalytic atoms during a reaction need not by itself lead to an inapplicable substrate rule. Consider, for example, the overlay graph in \hyperref[fig:transaminase]{Figure~\ref{fig:transaminase}A}, where the oxygen atom of a water molecule becomes an acyl group, and an acyl group of an educt is released as a water molecule. Since both hydrogen atoms of the water molecules are active participants in this reaction, they become part of the composite rule and there is no difficulty with including \ch{H2O} into the catalytic graph $C$ when constructing the substrate rule. Hence the issue is for the composite rule to include \enquote{enough} context. If the context is minimal, because the rules only refer to the reaction center as they do in the present paper, inapplicable substrate rules can arise. The modeler is, of course, at liberty to add more context to the composite rule so as to make the substrate rule applicable, but this then might be dictated more by mathematical than chemical necessity.

\begin{figure}[!ht]
\begin{center}
\includegraphics[scale=.8]{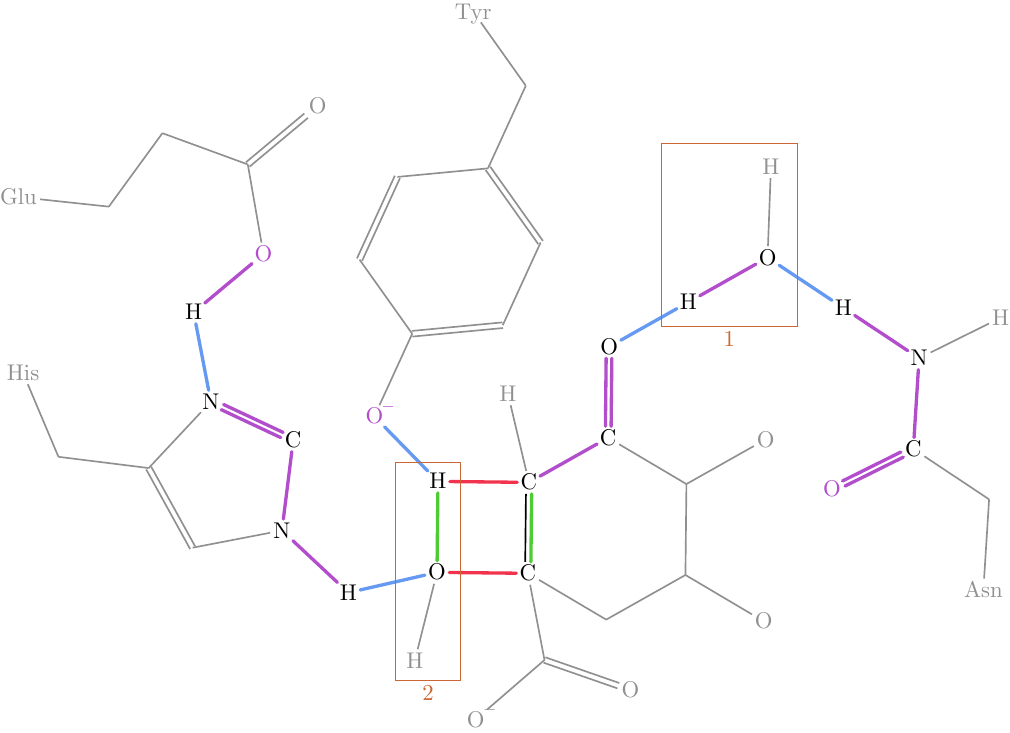}
\caption{%
An overlay graph of a 3-dehydroquinate dehydratase reaction (M-CSA \mcsaentry{55}). Two water molecules, framed by brown boxes, occur in the reaction. The water molecule labeled $1$ is catalytic, serving as a proton relay between the asparagine and the substrate. The other, labeled $2$, is a product formed by elimination of the alcohol group from the substrate.}
\label{fig:multi_water}
\end{center}
\end{figure}

Another interesting case is provided by M-CSA \mcsaentry{55}, which describes a reaction that dehydrates 3-dehydroquinate to 3-dehydroshikimate. An \ch{H2O} molecule must be supplied as an educt because it is needed as a catalyst in the first two steps of the reaction (where it twice establishes a hydrogen relay). A new \ch{H2O} molecule is created in this process by removing an \ch{OH} group from the substrate and combining it with a hydrogen from a histidine residue of the enzyme. At this point the mechanism has yielded a \ch{H2O} and 3-dehydroshikimate, in addition to the original \ch{H2O} molecule. However, the chemical state of the enzyme still needs to be restored. This is accomplished in the final step by using a \ch{H2O} molecule once more as a hydrogen relay. In terms of the mixture contents in that phase of the mechanism, the second catalytic deployment of water might involve the just newly generated \ch{H2O} molecule, rather than the one present initially. Overall, the reaction starts out with an \ch{H2O} and  3-dehydroquinate to yield \ch{2 H2O} and 3-dehydroshikimate. The overlay graph of the composite rule is shown in \hyperref[fig:multi_water]{Figure~\ref{fig:multi_water}}, in which we identify and label the water molecules.

When constructing the substrate rule, the question arises as to who are the catalysts in this reaction. Clearly, the enzyme is a catalyst, but so is the \ch{H2O} \#$1$ which catalyses the first two steps of the mechanism and can obviously also catalyze the third clean-up step; but so could be a combination of \ch{H2O} \#$1$ and \ch{H2O} \#$2$. When choosing the amino acid residues and \ch{H2O} \#$1$ as the catalytic graph $C$, we obtain an applicable substrate rule. In this case, both $c_1$ and $c_2$ identify the same \ch{H2O} \#$1$ on the left and the right of the rule, thus automatically maximizing the compatibility with the atom map of the mechanism. If $c_2$ were to point at \ch{H2O} \#$2$ instead, we would run into the same applicability issue discussed in the context of \hyperref[fig:isochorismate]{Figure~\ref{fig:isochorismate}B}.

That said, \ch{H2O} \#$2$ clearly can be a catalyst in the third step of the mechanism. From a network perspective, \ch{H2O} \#$2$ is therefore both a product and a catalyst. Indeed, it makes the reaction auto-catalytic (albeit in a likely uninteresting way, given the abundance of water). Yet, the substrate rule approach is unable to capture \ch{H2O} \#$2$ as a catalyst, since it does not appear on the educt side of the overall reaction.

}

\end{document}